\documentclass[a4paper,twoside,openright,11pt]{report}

\usepackage{amsbsy,latexsym}
\usepackage{graphicx}
\usepackage{fancyhdr}
\usepackage{amsfonts,amsmath,amssymb}

\newcommand{\id}{\mathbb{I}}
\newcommand{\im}{\mathrm{i}}
\newcommand{\tr}{{\rm tr}}
\newcommand{\ket}[1]{\vert #1 \rangle}

\newcommand{\braket}[2]{\langle #1 \vert #2 \rangle}
\newcommand{\ketbra}[2]{\vert #1 \rangle \langle #2 \vert}
\newcommand{\exvalue}[2]{\langle \hat{#1} \rangle_{#2}}
\newcommand{\sandwich}[3]{\langle #1 \vert \hat{#2} \vert #3 \rangle}
\newcommand{\comm}[2]{[\hat{#1},\hat{#2}]}
\newcommand{\acomm}[2]{\{\hat{#1},\hat{#2}\}}
\newcommand{\ppoison}[2]{\{\mathcal{#1},\mathcal{#2} \}_{pp}}

\def\lgem{\discretionary{l-}{l}{\hbox{l$\cdot$l}}}

\newtheorem{theorem}{Theorem}[section]
\newtheorem{lemma}{Lemma}[section]

\begin{document}

\pagenumbering{roman}

\thispagestyle{empty}

\begin{center}
 \includegraphics[height=3.5cm,width=5cm]{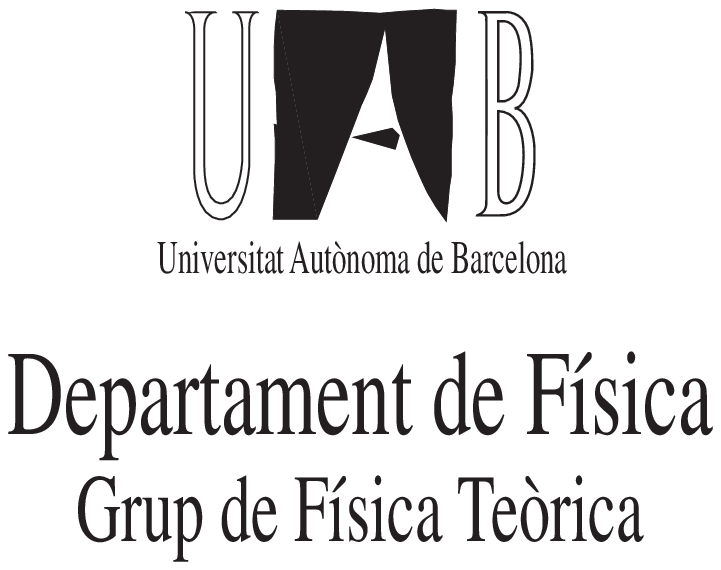}
\end{center}

\vspace{3cm}

\begin{center}
 {\LARGE \textbf{Efficiency in Quantum Key Distribution}}\\
 \vspace{0.2cm}
 {\LARGE \textbf{Protocols using entangled Gaussian states}}\\
 \vspace{1cm}
 {\Large Carles Rod\'o Sarr\'o}
\end{center}

\vspace{5cm}

{\large Treball de recerca del programa de Doctorat de F\'isica de la
Universitat Aut\`onoma de Barcelona, realitzat per Carles Rod\'o
Sarr\'o sota la direcci\'o de la Dra. Anna Sanpera i Trigueros.}

\begin{flushright}
 Bellaterra, Abril de 2008
\end{flushright}

\newpage

\vspace{3cm}

La Dra. Anna Sanpera i Trigueros, Investigadora ICREA al departament
de F\'isica Te\`orica de la Facultat de Ci\`encies de la Universitat
Aut\`onoma de Barcelona,\\
CERTIFICA: Que la present mem\`oria, que porta per t\'itol:
``Efficiency in Quantum Key Distribution'', ha estat realitzada sota la
seva direcci\'o pel Carles Rod\'o i Sarr\'o i que constitueix el seu
Treball de Recerca del programa de Doctorat de F\'isica.

\vspace{1cm}

\begin{flushright}
 Bellaterra, Abril de 2008
\end{flushright}

\vspace{1cm}

\begin{flushright}
 Anna Sanpera i Trigueros
\end{flushright}

\newpage

\vspace{3cm}

Vull donar els meus agra\"iments, en primer lloc a l'Albert Bramon
qui, far\`a ja uns anys, em va comen\c car mostrant els misteris
de la F\'isica Qu\`antica en aquelles tan ben cronometrades
classes i, qui, em va facilitar poder treballar amb l'Anna
Sanpera.

Vull agrair, com no, la inagotable paci\`encia i l'esfor\c c que ha
posat en mi l'Anna. No vull oblidar la generositat i lo molt que he
apr\`es co\lgem aborant amb el Gerardo Adesso.

Tamb\'e vull recordar molt especialment als meus ex-companys amb qui
m\'es temps he passat, el Javi i l'\`Alex per les variopintes
discusions de F\'isica i pels bons moments compartits fora de la feina.
Tamb\'e a les noves ``promeses'' Marc i Joan Antoni.

Als amics d'arreu, Rub\'en, Dani, \`Alex, Ignasi, Marina, Mara, Sito, Xavi,
Pepe, Cristina, Hugo, Edu, Jaime, Pedro, Shona, Mirka, Guillem,
Felix, Deivid, Chechu, Christian, ...

Finalment agrair a la fam\'ilia que sempre m'ha recolzat, al meu pare
per haver-me motivat a encuriosir-me per la F\'isica i sobretot a la
persona amb qui m\'es m'he barallat, la meva mare, per\`o qui ho ha
donat tot sempre per mi, i qui m\'es es mereix els meus agra\"iments.

\begin{flushright}
 Carles
\end{flushright}

\newpage

\
\newpage
\
\vspace{2cm}

\noindent {\Huge \textbf{Resum}}\\

\noindent L'estudi de l'efici\`encia en els protocols de
Criptografia \'es una tasca clau per a la seva futura
implementaci\'o tecnol\`ogica ja que els recursos f\'isics dels
quals es disposa experimentalment no s\'on i\lgem imitats. En
aquesta tesina s'estudia l'efici\`encia en els protocols de
distribuci\'o de claus qu\`antiques basats en l'entrella\c cament
qu\`antic. Usant estats de variable discreta i desti\lgem ant
entrella\c cament, sempre \'es possible desti\lgem ar singlets i
aix\'i extreure correlacions cl\`assiques perfectes. En canvi,
mitjan\c cant estats de variable cont\'inua com ho s\'on els modes
de la llum, la desti\lgem aci\'o no \'es possible mitjan\c cant
nom\'es operacions Gaussianes. No obstant, mostrem que, utilitzant
estats Gaussians de llum bipartits i l'entrella\c cament en les
quadratures del camp electromagn\`etic d'aquests estats, s\'i \'es
possible extreure seq\"u\`encies de bits correlacionats,
necess\`aries per desti\lgem ar claus aleat\`ories segures de
forma eficient. Obtenim que, usant estats mescla Gaussians NPPT
sim\`etrics i operacions Gaussianes es poden distribuir claus
qu\`antiques segures davant d'atacs individuals d'un espia. Davant
d'atacs finits coherents no tots els estats permeten una
extracci\'o segura. Mostrem quins estats s\'i s\'on segurs davant
d'atacs finits coherents i l'efici\`encia amb la qual es poden
extreure aquestes correlacions cl\`assiques davant d'atacs
individuals. Finalment s'analitza la relaci\'o entre
l'efici\`encia, l'entrella\c cament i la puresa per estats
Gaussians sim\`etrics. Concloem que un entrella\c cament gran
(m\'es correlacions entre els modes) implica un guany en
efici\`encia a l'hora d'extreure bits ben correlacionats. No hi ha
una relaci\'o un a un donat que estats amb el mateix entrella\c
cament per\`o diferents pureses donen lloc a diferent
efici\`encia. El que succeeix \'es que estats amb una puresa alta
afavoreixen un augment en l'efici\`encia donat que estem
desentrella\c cant l'espia, per\`o aquests estats s\'on
ineficients quan finalment realitzem Classical Advantatge
Distillation on, per extreure la clau aleat\`oria segura un alt
grau de correlacions \'es essencial.


\tableofcontents

\pagestyle{fancy}
\renewcommand{\chaptermark}[1]{\markboth{#1}{}}
\renewcommand{\sectionmark}[1]{\markright{\thesection\ #1}}
\fancyhf{} \fancyhead[LE,RO]{\bfseries\thepage}
\fancyhead[LO]{\bfseries\rightmark}
\fancyhead[RE]{\bfseries\leftmark}
\renewcommand{\headrulewidth}{0.5pt}
\renewcommand{\footrulewidth}{0pt}
\addtolength{\headheight}{0.5pt}
\fancypagestyle{plain}{\fancyhead{} \renewcommand{\headrulewidth}{0pt}}


\chapter{Introduction}

\pagenumbering{arabic}

Rolf Landauer, known for his remarkable contributions to the theory of
electrical conductivity, mesoscopic phenomena, and the physics of
information, during the 1960s, as the director of IBM Research's
Solid State Science Division persuaded the idea that information was
rather than an abstract concept a physical one.

Nowadays the Quantum Information science (QI) has become one of
the most active fields of research. QI research spreads in two
main areas, Quantum Communication and Quantum Computation tasks.
One of the most important topics of Quantum Communication deals
with the study of secure communications. Thus Cryptography, based
in protocols that exploit Quantum Mechanics properties, plays a
central role in Quantum Communication.

Cryptography refers to strategies which permit the secure
communication between two distant parties (traditionally denoted
by Alice and Bob) that wish to communicate secretly. So its
purpose is to design new communication algorithms being sure that
secrecy is preserved.
In Classical Cryptography there is only one cryptographic protocol,
known as the ideal Vernam cypher method, which is absolutely secure.
The Vernam cypher consists of a random secret key (private key) shared
between Alice (the sender) and Bob (the receiver) used to encode and
decode messages. However, this method
suffers from two drawbacks. First, if Alice wants to communicate
$N$ bits to Bob, they will need to have a random secret key in
advance with at least $N$ bits. Moreover, this key can only be
used once to make the method unbreakable. Second, the key must
originally be exchanged by hand before the communication to keep
secrecy and it is essential that the key is totally random. To
solve the second issue, classical ways exist to distribute the key
among them being this distribution partially secure. This problem
is known as Key Distribution.

In Classical Cryptography the problem of Key Distribution can be
solved partially by designing new algorithms which permit the
distribution of the secret key in a ``practical secure'' way. The
price to pay is that absolute security is not achievable. Nevertheless
as security relies on the fact that to decrypt one needs to invert
difficult mathematical operations, the time to decrypt the key is
long, and thus practical security is achievable.

The distribution problem is classically solved by using Public Key
Cryptosystems. They work as follows: Alice and Bob possess, in
advance, a common secret key. This secret key can be used many
times to distribute among them, several private keys that they are
going to use to encrypt messages through {\em e.g.} the Vernam
cypher method. Any time Alice (the sender) wants to distribute a
private key to encrypt later messages, she only needs to make one
key publicly available, referred as the public key. From this
public key any receiver can extract the private key, but only Bob
(the receiver in possesion of the secret key) can extract it in an
efficient way. Any receiver without the secret key needs to
decrypt a problem which has non-polynomial (NP) complexity. This
way of distributing a secure key can be done even when the key's
length is much smaller than the message and even if the key is
used several times.

Nowadays bank's security, electronic commerce and the internet are
mostly based on one of these algorithms, the RSA Cryptography algorithm
proposed in 1977 by {\em R}ivest, {\em S}hamir, and {\em A}dleman
from the MIT. Security in RSA algorithm relies on the fact that to
decrypt the algorithm one needs to solve the factorisation problem
which is an NP problem. Even now, there is no efficient algorithm
known to solve the factorisation problem. This means that even though
computational resources increase constantly, one simply needs to
exploit the NP character of the algorithm to make the solution harder
to find. Moreover, the key can eventually be redistributed and
changed.

The ``Quantum Computer`` arises here first as a menace for
Classical Key Distribution methods and then as a the solution for
the security in Cryptography. Based on the Quantum nature of the
microscopic world, this new generation of computers, still in a
theoretical stage, are known to be able to solve hard mathematical
problems rapidly. In 1994 Peter Shor proposed a Quantum
protocol to solve the factorisation problem in an efficient
way, known as the Shor's algorithm. If such a computer can be
realised, current cryptographic protocols will not be anymore secure.

Can Quantum Mechanics then offer a solution for a secure Cryptography
method? The answer is yes. Based on the intrinsic nature of the
microscopic world, Quantum Mechanics permits to perform Cryptography
in an unconditional secure way. At present, Quantum Cryptography is
the only real implementation of QI. In this work, we will present one
way of performing efficient Quantum Cryptography using entangled
Gaussian states of light and standard optical devices (Gaussian
operations).

\section{Classical Cryptography}

\subsection{Vernam cypher}

In Classical Cryptography there is a protocol called Vernam cypher
which is absolutely secure. This protocol is the best and most
well known classical private key (or ONE-TIME-PAD) cryptosystem.
In order to achieve security in communications between two parties
(Alice and Bob) with the Vernam cypher protocol, one needs to have
a private key, in advance, possessed by Alice and Bob exclusively
of at least the same length ($\#$ of bits) as the message to be
encoded. This key has to be random and in possession of Alice and
Bob only, and so a Classical Key Distribution is a pre-condition
to achieve this absolute security. If this is the case, then
Cryptography is going to be absolutely secure and works as follows.\\

If Alice wants to encode a message $m$ with the key $k$, she only
needs to perform the following operation between the message and
the key, and send the encoded message $e$ to Bob:
\begin{equation}
 {\rm Enc}_k(m)=m \oplus k=e.
\end{equation}
Only Bob who has also the key can decode the message (invert the
operation) because the key Alice has used is random. Thus, he only
needs to use the key again and perform the following operation to
the encoded message $e$ to retrieve the original message $m$:
\begin{equation}
 {\rm Dec}_k(e)={\rm Dec}_k [{\rm Enc}_k(m)] =e\oplus k=d=m.
\end{equation}
Let us illustrate Vernam cypher with an specific example. Alice
wants to communicate to Bob, in a secure way, a message $m$ (in a
binary string) of {\em e.g.} 9 bits. They share the key $k$ of the
same size as the message (9 bits). Alice encodes her message by
applying a $XOR$ (exclusive $OR$)~\footnote{Also known as $AND$ or
$\oplus \hspace{-1.5mm} \mod (2)$.} operation between the message
$m$ and the key $k$.

In the following table we summarise the $XOR$ operation

\begin{equation}
 \begin{tabular}{c|cc}
 $\oplus$ & 0 & 1\\
 \hline
 0 & 0 & 1\\
 1 & 1 & 0
 \end{tabular}
\end{equation}

As a result Alice has the encoded message $e$, that will send to
Bob in a public way.

\begin{equation}
 \begin{tabular}{c|l}
  message & $m=010011101\,$\\
  \hline
  key & $k=110100011$\\
  \hline
  \hline
  encoded message & $e=1000111110$\\
 \end{tabular}\nonumber
\end{equation}

Then Bob wants to readout the message, and thus, performs the
inverse operation (which is again a $XOR$) between the encoded
message $e$ and the key $k$.\\

As a result Bob has the decoded message $d$ that coincides with the
message Alice wanted to communicate to him.

\begin{equation}
 \begin{tabular}{c|l}
  encoded message & $e=1000111110\,$\\
  \hline
  key & $k=110100011$\\
  \hline
  \hline
  decoded message & $d=010011101$\\
 \end{tabular}\nonumber
\end{equation}

\subsection{Public key distribution: The RSA algorithm}

Thus far, Classical Cryptography has not solved the distribution
of the private key needed to perform Vernam cypher encryption.
This distribution can be done in a public and practical secure way
with the RSA algorithm. Nowadays the RSA algorithm is the most
commonly used algorithm of public key. It is a Classical Key
Distribution algorithm that permits together with the Vernam
cypher method to perform secure communication as long as the
factorisation problem is unsolved. Let us illustrate how it works
with the following example.

{\em i)} The sender, say Alice, chooses two "big" different prime
numbers, say $p=61$ and $q=53$ and computes its product
$n=p\,q=3233$ and also the following quantity
$\phi=(p-1)(q-1)=3120$.

{\em ii)} She chooses a positive integer $l$ smaller and coprime
with $\phi$, in the example $l=17$.

{\em iii)} As a private key, Alice gives to Bob the number $k$
such that $k \, l=1 \hspace{-1.5mm} \mod (\phi)$, take {\em e.g.}
$k=2753$. At the same time Alice makes public $l$ and $n$, what we
call public key.

{\em iv)} With the public key ($l$ and $n$) anyone can encrypt a
message $m$ and send it to Bob, but only Bob who is in possesion
of the private key $k$ is able to decrypt the message. This method
can thus be used to perform Classical Key Distribution. Any time
Alice wants to communicate with Bob, she sends a secure key
encrypted with $l$ and $n$ and only Bob will be able to retrieve
it. Once Bob has the secure key, Alice can send messages to Bob
{\em via} the Vernam cypher using this secure key.

{\em v)} Encryption proceeds as follows. Alice wants to distribute
a key encoded in a message $m=123$. She uses the public key and
computes the encryption ${\rm Enc}_{l,n}(m)=m^l
\hspace{-1.5mm}\mod(n)=e=855$.

{\em vi)} Bob now wants to decrypt the message $e$ to extract a
secure key, so he calculates ${\rm Dec}_{k,n}(e)={\rm Dec}_{k,n}
[{\rm Enc}_{l,n}(m)] =e^k \hspace{-1.5mm}\mod(n)=d=m=123$. Bob is
the only one in possesion of the private key $k$ and so the only
one that can decrypt the message $e$ to find the secure key Alice
is going to use with the Vernam cypher to communicate securely
with him.

The private key Alice and Bob share can be used more than once to
distribute secure keys, in such a way that, to perform Vernam
cypher we no longer need to share a long private key because we
can distribute many of them in a secure way. Security relies in
the fact that, from the encrypted message $e$, and the public keys
$l$ and $n$ it is very difficult to find the private key $k$ (and
so $m$) or the original two prime numbers $q$ and $p$, even in the
case we are reusing the key $k$. This is because factorisation is
a NP problem, whose efficient solution is not known yet. Security
thus, in the RSA algorithm, relies on the fact that with the
current computation resources, NP problems cannot be solved
efficiently.

\section{The solution to the distribution of the key}

What Quantum Cryptography offers is an absolutely secure
distribution of a random key which combined with the Vernam cypher
guarantees completely secure Cryptography. Thus, the Quantum
Cryptography problem is in fact the problem of distributing a
secure random key, {\em i.e.} the Quantum Key Distribution (QKD)
problem.

Quantum Cryptography relies on the possibility of establishing a
secret random key between two distant parties traditionally
denoted by Alice and Bob. If the key is securely distributed, the
algorithms used to encode and decode any message can be made
public without compromising security. The key consists typically
in a random sequence of bits which both, Alice and Bob, share as a
string of classically correlated data. The superiority of Quantum
Cryptography comes from the fact that the laws of Quantum
Mechanics permit to the legitimate users (Alice and Bob) to infer
if an eavesdropper has monitored the distribution of the key and
has gained information about it. If this is the case, Alice and
Bob will both agree in withdrawing the key and will start the
distribution of a new one. In contrast, Classical Key
Distribution, no matter how difficult the distribution from a
technological point of view is, can always be intercepted by an
eavesdropper without Alice and Bob realising it.

In Quantum Cryptography two seemingly independent main schemes
exist for QKD. The first, the ''Prepare and Measure`` scheme,
originally proposed by C.H. Bennett and G. Brassard in 1988 and
known as BB84 \cite{BB84}, does not use entangled states shared
between Alice and Bob and the key is established by sending
non-entangled quantum states between the parties and communicating
classically. Security is guaranteed by the Quantum nature of the
measurements. The second scheme (''Entanglement based``), uses as
a resource shared entanglement, like the one originally proposed
by A. Ekert in 1991 known as Ekert91 \cite{Ekert91}, where indeed
entanglement is explicitly distributed and the security is
guaranteed by Bell's theorem. However, the two schemes have been
shown to be completely equivalent \cite{BB92}, and specifically
entanglement stands as a precondition for any secure key
distribution \cite{Curty04}.

\subsection{''Prepare and Measure`` scheme}

BB84 permits a secure distribution of a secret key. The protocol
does not avoid an eavesdropper from intercepting the key, but it
lets Alice and Bob know if the key has been intercepted, and so
that it can be discarded. Security relies in the Quantum nature of
the measurement. We sketch here the steps of the protocol.

{\em i)} Alice prepares a secret sequence of random bits and
encodes them in the state of a spin-1/2 system (or the
polarisation of photons) by choosing randomly between two bases (Z
and X). Alice encodes $\ket \pm$ ($\ket \pm_x$) according to $0/1$
in base Z ($0/1$ in base X). Then she sends Bob the states she has
prepared. As an example:

\begin{equation}
 \begin{tabular}{c|ccccccccc}
 Alice random bits & 0 & 1 & 1 & 0 & 0 & 1 & 1 & 0 & 0\\
 \hline
 Alice random bases & Z & X & X & X & Z & X & Z & X & X\\
 \hline
 Alice
 final states & $\ket{+}$ & $\ket{-}_x$ & $\ket{-}_x$ & $\ket{+}_x$
 & $\ket{+}$ & $\ket{-}_x$ & $\ket{-}$ & $\ket{+}_x$ & $\ket{+}_x$\\
 \end{tabular}\nonumber
\end{equation}

{\em ii)} Bob receives the states and measures in another random
choice of bases. The outcome of the measurements is going to be
retained as the bits received.

\begin{equation}
 \begin{tabular}{c|ccccccccc}
 Alice final states & $\ket{+}$ & $\ket{-}_x$ & $\ket{-}_x$ &
 $\ket{+}_x$ & $\ket{+}$ & $\ket{-}_x$ & $\ket{-}$ & $\ket{+}_x$
 & $\ket{+}_x$\\
 \hline
 Bob random bases & X & Z & X & X & X & X & X & X & X\\
 \hline
 Bob final
 states & $\ket{+}_x$ & $\ket{+}$ & $\ket{-}_x$ & $\ket{+}_x$ &
 $\ket{+}_x$ & $\ket{-}_x$ & $\ket{-}_x$ & $\ket{+}_x$ &
 $\ket{+}_x$\\
 \hline
 Bob received bits & 0 & 0 & 1 & 0 & 0 & 1 & 1 & 0 & 0\\
 \end{tabular}\nonumber
\end{equation}

{\em iii)} Bob communicates to Alice his choice of basis in a public
way.

{\em iv)} Alice identifies the set of bits for which they have
performed the measurement in the same basis, {\em i.e.} outcomes
3, 4, 6, 8 and 9. Alice and Bob discard the set of data in which
they did not agree (the rest).

{\em v)} Bob sends part of his data (received bits) to Alice by a
public channel. Alice checks the correlation between the data.

{\em vi)} If the error rate is less than 25$\%$ Alice deduces that
there is not an eavesdropper present and she communicates it to
Bob. Alice and Bob use the set of remaining data as a private key
to encrypt messages with Vernam cypher.

\subsection{''Entanglement based`` scheme}

Other protocols exist, which demand as a fundamental resource,
shared entanglement between Alice and Bob. In the same way as in
BB84, these protocols permit a secure distribution of a secret key.
This can be done as far as the protocol ensures if there has been
an interception of the key. Security relies in the Quantum
correlations {\em i.e.} entanglement like in Ekert91. We sketch
here the steps of this well-known protocol below.

{\em i)} The first step consists on distributing (along the $z$
direction) singlet states of a spin-1/2 system (or polarisations
of photons) between Alice and Bob. Thus Alice and Bob share many
copies of a Bell state $\ket {\Psi^-}=\frac{1}{\sqrt 2}(\ket 0
\ket 1-\ket 1 \ket 0)$.

{\em ii)} Alice and Bob are going to measure in the $x-y$ plane in
one of the three directions given by unit vectors $\vec A_i =
(\cos \phi_i^A, \sin \phi_i^A)$ and $\vec B_j = (\cos \phi_j^B,
\sin \phi_j^B)$ respectively, where the azimuthal angles are fixed
to $\phi_i^A=(0,\pi/4,\pi/2)^i$ and to
$\phi_j^B=(\pi/4,\pi/2,3\pi/4)^j$. Each time they will choose the
basis randomly and independently for each pair of incoming
particles.

{\em iii)} The quantity
\begin{equation}
 \mathcal E(\vec A_i,\vec B_j) = \mathcal P_{++}(\vec A_i,\vec
 B_j) + \mathcal P_{--}(\vec A_i,\vec B_j) - \mathcal P_{+-}(\vec
 A_i,\vec B_j) - \mathcal P_{-+}(\vec A_i,\vec B_j)
\end{equation}
is the correlation coefficient of the measurements performed by
Alice and Bob along $\vec A_i$ and by Bob  along $\vec B_j$. Here
$\mathcal P_{\pm \pm}(\vec A_i,\vec B_j)$ denotes the probability
that result $\pm1$ has been obtained along $\vec A_i$ and  $\pm1$
along $\vec B_j$. Straightforward calculations give rise to

$\mathcal P_{++}(\vec A_i,\vec B_j) = \frac{1}{2} \sin^2(\phi_i^A-\phi_j^B)$,

$\mathcal P_{--}(\vec A_i,\vec B_j) = \frac{1}{2} \sin^2(\phi_i^A-\phi_j^B)$,

$\mathcal P_{+-}(\vec A_i,\vec B_j) = \frac{1}{2} \cos^2(\phi_i^A-\phi_j^B)$,

$\mathcal P_{-+}(\vec A_i,\vec B_j) = \frac{1}{2} \cos^2(\phi_i^A-\phi_j^B)$,

thus according to the Quantum rules $\mathcal E(\vec A_i,\vec B_j)
= - \vec A_i \vec B_j = \cos(\phi_i^A-\phi_j^B)$. We see that
whenever they choose the same orientation Quantum Mechanics
predicts total anticorrelation in the outcomes {\em i.e.}
$\mathcal E(\vec A_i,\vec B_j)=-1$.

{\em iv)} Here we define a quantity composed of those correlation
coefficients for which Alice and Bob have measured in different
directions,
\begin{equation}
 \mathcal S = |\mathcal E(\vec A_1,\vec B_1) + \mathcal E(\vec
 A_3,\vec B_3) - \mathcal E(\vec A_1,\vec B_3) + \mathcal E(\vec
 A_3,\vec B_1)|.
\end{equation}
Again, Quantum Mechanics requires, $\mathcal S = 2\sqrt{2} > 2$.

{\em v)} After the transmission has taken place, Alice and Bob can
announce in public the orientations they have chosen for each
measurements and divide them into two separated groups. A first
group for which they coincide and a second group for which they do
not. The second group of outcomes is made public and it is used to
establish the value of $\mathcal S$.

The CHSH (Clauser, Horne, Shimony, and Hold) inequalities, a
generalisation of Bell inequalities, asserts that $\mathcal S \leq
2$. But Quantum Mechanics, and in particular Bell states, violate
CHSH inequalities. If this is the case, {\em i.e.} the value of
$\mathcal S$ that they find is exactly $ 2\sqrt{2}$, they now that
their states have not been disturbed and so the first group of
outcomes, that are random, are totally anticorrelated and can be
converted into a secret string of bits. Later on they can use this
string as a private key to encrypt messages with Vernam cypher.

\section{Efficiency in Quantum Key Distribution (QKD) protocols}

As already mentioned, Quantum Key Distribution, refers to specific
Quantum strategies which permit the secure distribution of a
secret key between two parties that wish to communicate secretly.
Quantum Cryptography has been proved unconditionally secure in
ideal scenarios and has been successfully implemented using
quantum states with finite (discrete) as well as infinite
(continuous) degrees of freedom. We have analysed the efficiency
of QKD protocols that use as a resource entangled Gaussian states
and Gaussian operations only. In this framework, it has already
been shown that QKD is possible \cite{Navascues05} but the issue
of its efficiency was not considered. We have propose a figure of
merit (the efficiency $E$) to quantify the number of classical
correlated bits that can be used to distill a key from a sample of
$N$ entangled states. We have related the efficiency of the
protocol to the entanglement and purity of the states shared
between the parties.

Notice that if Alice and Bob share a collection of distillable
entangled states, they can always obtain a smaller number of
maximally entangled states from which they can establish a secure
key \cite{Deutsch96}. The number of singlets (maximally entangled
states) that can be extracted from a quantum state using only
Local Operations and Classical Communication (LOCC) is referred to
as the Entanglement of Distillation $E_D$. In order to establish a
key, another important concept is the number of secret bits $K_D$,
that can be extracted from a quantum state using LOCC. As a secret
bit can always be extracted from maximally entangled state, $E_D
\leq K_D$. Furthermore, there are quantum states which cannot be
distilled but in spite of being entangled, {\em i.e.}, have
$E_D=0$. They are usually referred to as bound entangled states
since its entanglement is bound to the state. Nevertheless, for
some of those states it has been shown that $K_D \neq 0$ and thus
they can be used to establish a secret key \cite{Horodecki05}.

A particular case of states that cannot be ``distilled'' by
``normal'' procedures are continuous variables Gaussian states,
{\em e.g.}, coherent, squeezed and thermal states of light. By
``normal'' procedures we mean operations that preserve the
Gaussian character of the state (Gaussian operations). They
correspond {\em e.g.} to beam splitters, phase shifts, mirrors,
squeezers, etc. Thus, in the Gaussian scenario all entangled
Gaussian states posses bound entanglement. Quantum Cryptography
with Gaussian states using Gaussian operations has been
experimentally implemented using ``Prepare and Measure'' schemes
with either squeezed or coherent states \cite{Preskill01,
Grangier02, Silberhorn02}. Those schemes do not demand
entanglement between the parties.

Navascu{\'e}s {\em et al.} \cite{Navascues05} have shown that it
is also possible using only Gaussian operations to extract a
secret key {\em {\`a} la} Ekert91 from entangled Gaussian states,
in spite the fact that these states are not distillable. In other
words, it has been proven that in the Gaussian scenario all
entangled Gaussian states fulfil $GK_D>0$ (where the letter $G$
stands for Gaussian) while $GE_D=0$.

The way to proceed in this scheme {\em i.e.}, how can, Alice and
Bob extract a list of classically correlated bits from a set a of
symmetric $1 \times 1$ entangled modes goes as follows:

i) they agree on a value $x_0 > 0$,

ii) Alice(Bob) measures the quadrature of each of her(his) modes
$\hat X_A(\hat X_B)$,

iii) they make public the modulus of their outcomes, but not the
sign and accept only outputs such that $|x_A|=|x_B|=x_0$,

iv) they associate {\em e.g.}, the classical
value $0$(1) to $x_i=+x_0(-x_0)$, $i=A, B$ and thus establish a
list of classically correlated bits.

From there, they can apply Classical Advantage Distillation
\cite{Maurer93} to establish the secret key. This protocol is
secure against individual eavesdropper attacks. As the protocol is
based on output coincidences of the measurements of the
quadratures which, by definition, are operators with a continuous
spectrum, this protocol has zero efficiency.

\section{Outline}

Efficiency is a key issue for any experimental implementation of
Quantum Cryptography since available resources are not unlimited.
Since it is possible to extract a secret key {\em \`a la} Eckert91
from entangled Gaussian states in the Gaussian scenario one
important question is to address the efficiency.

In fact, the protocol suffers from an efficiency problem because
the success probability of the protocol is vanishingly small. Here
we study the consequences of relaxing the conditions to a more
realistic scenario. We assume that Alice and Bob can extract a
list of sufficiently correlated classical bits obtained by
accepting measurement outputs that do not coincide but are bound
within a range. We ask ourselves which is the possibility that
Alice and Bob can still distribute the key in a secure way under
individual and finite coherent attacks.

We find that there always exists a finite interval which the protocol
can be implemented successfully. The length of this interval depends
on the entanglement and on the purity of the shared states, and
increases with increasing entanglement.

In Chapter 2 we review the formalism of Continuous Variable (CV)
systems focussing on Gaussian states. It will be shown that Gaussian
states admit an easy mathematical description based on phase space
Wigner functions. We will introduce also the basic ingredients to
describe also systems with entanglement.

In Chapter 3 we will present first a simple academic protocol that
permits to extract a quantum key from an entangled continuous variable
system. Differently from discrete systems, Gaussian entangled states
cannot be distilled with Gaussian operations. However, as we will
show, it is possible to extract perfectly correlated classical bits
to establish a secret key between the sender and receiver. We will
then demonstrate that this protocol, properly modified, can be made
efficient and can be implemented with present technology \cite{Rodo07}.

Finally, in Chapter 4, we conclude summarising our results and giving
some general conclusions and future directions of research.

\chapter{Continuous Variable formalism}

This Chapter is intended exclusively to describe Continuous Variable
systems and to provide the mathematical framework to analyse the
problem of QKD in Continuous Variable systems. We will focus then on
Gaussian Continuous Variable states, which describe among others,
coherent, squeezed and thermal states of light. Presently these
states are the preferred resources in experiments of QI using
Continuous Variable systems. For further background information the
interested reader is referred to \cite{Braunstein05,vanLoock02,
Zhang90,Lee95}.

\section{Continuous Variable system}

A system corresponds to a continuous variable system if it
possesses two canonical conjugated degrees of freedom {\em i.e.}
there exist two observables that fulfil the Canonical Commutation
Relations (CCR). The CCR for two canonical observables $\hat q$
and $\hat p$ read~\footnote{Quadrature operators are chosen
dimensionless in such a way that $\hbar$ is not going to appear in
any formula.}

\begin{equation}\label{1CCR}
 \comm{q}{p} = \im \id.
\end{equation}
It is a direct consequence that they possess a continuous spectra and
act in an infinite dimensional Hilbert space.

As examples of CV systems we can think as the position-momentum of
a massive particle, the quadratures of an electromagnetic field or
the collective spin of a polarised ensemble of atoms. In all of the
three examples above there exist two observables fulfilling
\eqref{1CCR}. As we will show, they obey the standard bosonic
commutation relations and so we call these systems bosonic
modes. We can deal with several modes, and in this case ordering
the operators by canonical pairs by $\hat R^T = (\hat q_1,\hat
p_1,\hat q_2,\hat p_2,...,\hat q_N,\hat p_N)$ we can compactly
state CCR as

\begin{equation}\label{CCR}
 \comm{R_i}{R_j} = \im (J_N)_{ij}
\end{equation}
where $i,j=1,2,...,2N$ and $J_N = \oplus_{i=1}^N J$ accounts for
all modes while $J$ is the so-called symplectic matrix which
corresponds to a antisymmetric and non-degenerate form fullfilling
(i) $\forall \eta, \zeta\in \mathbb{R}^{2N}: \langle \eta \vert J
\vert \zeta \rangle = - \langle \zeta \vert J \vert \eta \rangle$
and (ii) $\forall \eta: \langle \eta \vert J \vert \zeta \rangle =
0 \Rightarrow \zeta=0$. In the appropriate choice of basis
(canonical coordinates) the symplectic matrix is brought into the
standard form $J=\begin{pmatrix}
    0 & 1 \cr
    -1 & 0
\end{pmatrix}$

\section{Canonical Commutation Relations}

The CCR \eqref{1CCR} are related with the classical Poisson brackets
{\em via} the $1^{\textrm{st}}$ quantisation transcription:
$\ppoison{A}{B} \equiv \sum_\mu (\frac{\partial \mathcal
A}{\partial Q_\mu} \frac{\partial \mathcal B}{\partial P_\mu} -
\frac{\partial \mathcal B}{\partial Q_\mu} \frac{\partial \mathcal
A}{\partial P_\mu}) \longrightarrow - \im \comm{A}{B} \equiv
-\im(\hat A \hat B - \hat B \hat A)$ and $\mathcal A
\longrightarrow \hat A$. They can be also described by using the
annihilation and creation operators $\hat a_\mu$ and $\hat
a_\mu^\dag$ which obey the standard bosonic commutation relations
\begin{equation}\label{ACCR}
 [\hat{a}_\mu,\hat{a}_\nu^\dag] = \delta_{\mu \nu}, \quad
 [\hat{a}_\mu,\hat{a}_\nu] = [\hat{a}_\mu^\dag,\hat{a}_\nu^\dag]=0
\end{equation}
$\mu, \nu = 1,2,...,N$. The CCR in forms \eqref{CCR} and \eqref{ACCR}
are related by a unitary matrix $U=1/\sqrt{2}\begin{pmatrix}
    \id_N & \im \id_N \cr
    \id_N & - \im_N  \id
\end{pmatrix}$ such that if we define $\hat
O^T = (\hat a_1,\hat a_2,...,\hat a_N,\hat a_1^\dag,\hat
a_2^\dag,...,\hat a_N^\dag)$ then $\hat O_i = U_{ij} \hat R_j$ .

The representation of the CCR up to unitaries is not unique. For
instance, for a single mode in the Schr\"odinger representation
each degree of freedom is embedded in $\mathcal{H}=\mathcal
L^2(\mathbb{R})$, while the operators $\hat q$ and $\hat p$ act
multiplicative and derivative respectively
\begin{equation}
 \left. \begin{array}{lcc}
 \hat q &=& q\\
 \hat p &=& -\im \frac{\partial}{\partial q}
 \end{array} \right\}
\end{equation}
but also $\hat q = +\im \frac{\partial}{\partial p}, \quad \hat p
= p$ is equally possible. In both representations the operators
are unbounded.

A way to remove ambiguities (up to unitaries) and to treat with
bounded operators is using the Weyl operators. The Weyl
operator is defined as
\begin{equation}
 \hat{W}_\zeta \equiv e^{\im \zeta^T \cdot J \cdot \hat{R}}
\end{equation}
where $\zeta^T = (\zeta_1, \zeta_2, ..., \zeta_{2N})$. Then the
Weyl operators satisfies the Weyl relation
\begin{equation}
 \hat{W}_\zeta \hat{W}_\eta = e^{-\frac{\im}{2}
 \zeta^T \cdot J \cdot \eta} \hat{W}_{\zeta+ \eta},
\end{equation}
or in an analogous way
\begin{equation}
 \hat{W}_\zeta \hat{W}_\eta = \hat{W}_\eta \hat{W}_\zeta e^{-\im
 \zeta^T \cdot J \cdot \eta }.
\end{equation}
\begin{theorem}
 \emph{(Stone-von Neumann theorem)}
 \label{svonn}
 Let $\hat{W}_1$ and $\hat{W}_2$ be two Weyl systems over a finite
 dimensional phase space $(N < \infty)$. If the two Weyl systems are
 strongly continuous~\footnote{$\forall \ket{\psi} \in  \mathcal{H}:
 \lim_{\zeta \to 0} ||\, \ket{\psi} - \hat{W}_\zeta \ket{\psi}|| = 0$
 .} and irreducible~\footnote{$\forall \zeta \in \mathbb{R}^{2N}:
 [\hat{W}_\zeta, \hat A] = 0 \Rightarrow \hat A \propto \id$.} then
 they are equivalent (up to an unitary).
\end{theorem}
According to the theorem $\ref{svonn}$ there exists only one
equivalent representation of the Weyl relation.

The Weyl operator acts in the states as a translation in the phase
space (displacements $e^{\im \eta \hat{p}} \ket{q} = \ket{q -
\eta}$ and kicks $e^{\im \zeta \hat{q}} \ket{q} = e^{\im \zeta q}
\ket{q}$) as it can be checked by looking to its action onto an
arbitrary position-momentum operator
\begin{equation}
 \hat{W}_\zeta^\dag \hat{R_i} \hat{W}_\zeta = \hat{R_i} - \zeta_i \id
\end{equation}

\section{Phase-space}

Phase space formulation of Quantum Mechanics offers a framework in
which Quantum phenomena can be described using as much classical
language as allowed. It appeals naturally to one's intuition and
can often provide useful physical insights. Furthermore, it
requires dealing only with constant number equations and not with
operators, which can be of significant practical advantage. This
mathematical advantage arises here from the fact that the
infinite-dimensional complex Hilbert space structure which is in
principle a difficult object to work with, can be mapped into the
linear algebra structure of the finite-dimensional real phase
space. We will extend this map (\ref{psgeometry}) and how to
characterise states and operations in sections (\ref{dfunctions})
and (\ref{soperations}) respectively.

\subsection{Phase space geometry}\label{psgeometry}

A system of $N$ canonical degrees of freedom is described
classically in a $2N$-dimensional real vector space~\footnote{They
are isomorphic (there exist a bijective morphism between the two
groups).} $V \simeq \mathbb{R}^{2N}$. Together with the symplectic
form it defines a symplectic real vector space (the phase space)
$\Omega \simeq \mathbb{R}^{2N}$. The phase space is naturally
equipped with a complex structure and can be identified with a
complex Hilbert space $\mathcal{H}_\Omega \simeq \mathbb{C}^{N}$.
If $\braket{\,\,}{\,\,}$ stands for the scalar product in
$\mathcal{H}_\Omega$ and $\braket{\,\,}{\,\,}_J$ for the
symplectic scalar product in $V$ their connection reads
\begin{equation}
 \braket{\eta}{\zeta} = \braket{J \eta}{\zeta}_J + \im
 \braket{\eta}{\zeta}_J
\end{equation}
where $\eta = (q,p) \in V$ while $\eta = q + \im p \in
\mathcal{H}_\Omega$ such that any orthonormal basis in
$\mathcal{H}_\Omega$ leads to a canonical basis in $V$. Moreover,
any unitary operator (which preserve the scalar product) acting on
$\mathcal{H}_\Omega$ leads to a symplectic operation $S$ in the
phase space in such a way that the symplectic scalar product is
also preserved. The inverse is also true provided that the
symplectic operation commutes with the symplectic matrix $J$.

\subsection{Symplectic operations}\label{soperations}

We hence define symplectic canonical transformation $S$ as these
ones which preserve the CCR and therefore leave the basic
kinematic rules unchanged. That is, if we transform our quadrature
operators $\hat R_S = S \cdot \hat R$ still equation \eqref{CCR}
is fulfilled. In a totally equivalent way we can define symplectic
transformation as the ones which preserve the symplectic scalar
product and therefore~\footnote{From now on we neglect the
subscript $N$ in symplectic matrix.}
\begin{equation}
 S^T \cdot J \cdot S = J.
\end{equation}
The set of real $2N\times2N$ matrices $S$ satisfying the above
condition form the symplectic group $Sp(2N,\mathbb{R})$. To
construct the affine symplectic group we just need to add also the
phase space translations whose group generators are $\hat
G^{(0)}_i = J_{ij}\hat R_j$. Apart from that, the group
generators of the representation of $Sp(2N,\mathbb{R})$ which
physically corresponds to the Hamiltonians which perform the
symplectic transformations on the states are of the form
$\frac{1}{2}\{ \hat R_i,\hat R_j \}$. This corresponds to
hermitian Hamiltonians of quadratic order in the canonical
operators. When rewriting them in terms of creation / annihilation
operators we can divide it into two groups.

Compact generators (passive):\\\\ $\hat G^{(1)}_{\mu \nu} = \im
\frac{(\hat a^\dag_\mu\hat a_\nu-\hat a^\dag_\nu\hat a_\mu)}{2}$,
\quad {$\hat G^{(2)}_{\mu \nu} = \frac{(\hat a^\dag_\mu\hat
a_\nu+\hat a^\dag_\nu\hat a_\mu)}{2}$\\\\
and non-Compact generators (active):\\\\ $\hat G^{(3)}_{\mu \nu} = \im
\frac{(\hat a^\dag_\mu\hat a^\dag_\nu-\hat a_\nu\hat a_\mu)}{2}$,
\quad $\hat G^{(4)}_{\mu \nu} = \frac{(\hat a^\dag_\mu\hat
a^\dag_\nu+\hat a_\nu\hat a_\mu)}{2}$.\\

The passive ones are generators which commute with all the number
operators $\hat n_\mu \equiv \hat a_\mu^\dag \hat a_\mu$ and so
they preserve the total number, in this sense they are passive. In
case our system corresponds to modes of the electromagnetic field,
then, what is being preserved is the total number of photons.
They can be implemented optically by only using beam splitters,
phase shifts and mirrors and conversely, only using them we can
implement any Hamiltonian constructed by a linear combination of
the compact generators. With all the generators we then have
enough to generate all the unitaries, $\hat U_\lambda = e^{\im \lambda
\cdot \hat G}$.

As the simplest example for, one mode ($N=1$) we have the phase
shift operator
\begin{equation}
 \hat U_\theta = e^{\im \theta \hat a^{\dag} \hat a}
\end{equation}
which amounts to the symplectic operation in phase space
\begin{equation}
 S_\theta = \begin{pmatrix}
        \cos{\theta} &  -\sin{\theta} \cr
        \sin{\theta} & \cos{\theta}
       \end{pmatrix}.
\end{equation}
On the other hand we have the active ones, they change the energy
of the state. The most important one is the single mode squeezing
operator ($N=1$), whose unitary expression for a fixed angle
$\phi=0$ and squeezing parameter $r>0$ reads
\begin{equation}
 \hat U_r = e^{\frac{r}{2}(\hat a^2 - \hat a^{\dag 2})}
\end{equation}
while its symplectic operation in phase space reads
\begin{equation}
 S_r = \begin{pmatrix}
        e^{-r} & 0 \cr
        0 & e^r
       \end{pmatrix}.
\end{equation}
Finally phase space translations for one mode whose unitary is
$\hat U_\alpha = e^{\alpha \hat a^\dag - \alpha^* \hat a}$ where
$\alpha = \frac{(q_0 + \im p_0)}{\sqrt 2}$, amounts to the
symplectic operation in phase space as $s_\alpha = \begin{pmatrix}
        q_0 \cr
        p_0
\end{pmatrix}$.
The most important non-trivial unitaries for two modes ($N=2$) are
beam splitters $\hat U_{BS} = e^{\frac{\theta}{2}(\hat a_1 \hat
a_2^\dag - \hat a_1^\dag \hat a_2)}$ (at fixed angle $\phi=0$,
reflectivity  $R=\sin^2{\theta/2}$ and transmitivity
$T=\cos^2{\theta/2}$) and two mode squeezings $\hat U_{TMS} =
e^{r(\hat a_1 \hat a_2 - \hat a_1^\dag \hat a_2^\dag)}$ that
amounts to
\begin{equation}
 S_{BS} = \begin{pmatrix}
        \cos{\theta/2} & 0 & \sin{\theta/2} & 0\cr
        0 & \cos{\theta/2} & 0 & \sin{\theta/2}\cr
    -\sin{\theta/2} & 0 & \cos{\theta/2} & 0\cr
    0 & -\sin{\theta/2} & 0 & \cos{\theta/2}
        \end{pmatrix}
\end{equation}
and
\begin{equation}
  \quad
 S_{TMS} = \begin{pmatrix}
        \cosh r & 0 & \sinh r & 0\cr
        0 & \cosh r & 0 & -\sinh r\cr
    \sinh r & 0 & \cosh r & 0\cr
    0 & -\sinh r & 0 & \cosh r
        \end{pmatrix}.
\end{equation}

\section{Probability Distribution functions}\label{dfunctions}

One of the most important tools of the phase space formulation of
Quantum Mechanics are the phase space probability distribution
functions. The best known and widely used is the Wigner
distribution function, but there is not a unique way of defining a
quantum phase space distribution function. In fact, several
distribution functions with different properties, rules of
association and operator ordering can also be well defined. For
instance sometimes normal ordered (P-function), antinormal ordered
(Q-function), generalised antinormal ordered (Husimi-function),...
distributions can be more convenient depending on the problem
being considered. In this dissertation we are only going to work
with the totally symmetrical ordered (Weyl ordered) one, the
Wigner distribution function.

Due to the fact that a joint probability at a fixed position $\hat
q$ and momentum $\hat p$ point is not allowed in Quantum Mechanics
by Heisenberg uncertainty theorem, the quantum phase space
distribution function should, therefore, be considered simply as a
mathematical tool that facilitates quantum calculations. Joint
probabilities can be negative, so that one deals with
quasiprobability distributions as long as it yields a correct
description of physically observable quantities.

\subsection{Quantum states}

At the level of density operators, $\hat \rho$ defines a quantum
state iff it satisfies the following properties
\begin{equation}
 \tr{\hat \rho} = 1, \quad \hat \rho \geq 0 \quad [\Rightarrow
 \quad \hat \rho^\dag = \hat \rho].
\end{equation}
This operator belongs to the bounded linear operators Hilbert space
$\mathcal B(\mathcal H)$. For pure states the ket and wave
function formulation is enough to describe our state. There, a
state lives in a Hilbert space $\mathbb{C}^d$ for qu{\em d}its, (a
system of discrete variables) or $ \mathcal L^2(\mathbb{R}^N)$ for
$N$ modes, (a system of continuous variables).

For systems of continuous variables, the Wigner distribution
function gives a complete description of the state. Given a state
$\hat \rho$ (a single mode) we define the Wigner distribution
function as~\footnote{For pure states the definition gets
simplified to $\mathcal{W}_\rho(q,p) = \frac{1}{\pi} \int dx \,
e^{-2\im px} \psi^*(q-x) \psi(q+x)$.}
\begin{equation}\label{wigner}
 \mathcal{W}_{\rho}(q,p) = \frac{1}{\pi} \int dx
 \sandwich{q+x}{\rho}{q-x} e^{-2\im px}.
\end{equation}
This transformation is called Weyl-Fourier transformation and it
gives the bridge between density operators and distribution
functions. Sometimes, for computational reasons it is better to
compute first the characteristic distribution function which is
obtained through
\begin{equation}\label{characteristic}
 \chi_\rho(\zeta, \eta) = \tr \{\hat{\rho}\hat{W}_{(\zeta, \eta)} \}.
\end{equation}
The above two distribution functions are fully equivalent in the
sense of describing completely our quantum state and are related
by a Symplectic-Fourier transform
\begin{equation}
 \mathcal{W}_\rho(q,p) = \frac{1}{(2 \pi)^2} \int d\zeta \int d\eta \,
 \chi_\rho(\zeta, \eta) e^{-\im \zeta p + \im \eta q} = \frac{1}{2\pi}
 \mathcal {SFT} \{ \chi_\rho(\zeta, \eta) \},
\end{equation}
\begin{equation}
 \chi_\rho(\zeta, \eta) = \int dq \int dp \, \mathcal{W}_\rho(q,p)
 e^{\im  \zeta p - \im  \eta q} = 2\pi \mathcal {SFT}^{-1} \{
 \mathcal{W}_\rho(q,p) \}.
\end{equation}
The Weyl-Fourier transformation is invertible and it provides a
way to recover our density operator from both distribution
functions
\begin{equation}
 \begin{split}
 \hat{\rho} &= \frac{1}{2 \pi} \int dq \int dp
 \int d\zeta \int d\eta \, \mathcal{W}_\rho(q,p) e^{-\im \zeta p + \im
 \eta q} \hat{W}_{(-\zeta, -\eta)} =\\
 &= \frac{1}{2 \pi} \int d\zeta \int d\eta \, \chi_\rho(\zeta,\eta)
 \hat{W}_{(-\zeta, -\eta)}.
 \end{split}
\end{equation}
At this level, $\mathcal{W}$ (and $\chi$) defines a quatum state
iff they satisfy the following properties
\begin{equation}
 \int dq \int dp \, \mathcal{W}(q,p) = 1, \quad \int dq \int dp \,
 \mathcal{W}(q,p) \mathcal{W}_p(q,p) \geq 0 \quad [\Rightarrow
 \quad \mathcal{W}^*(q,p) = \mathcal{W}(q,p)]
\end{equation}
for all pure states $\mathcal{W}_p$ and
\begin{equation}
 \chi(0,0) = 1, \quad \sum_{i,j=1}^{2N}a_i^*a_j \chi(\zeta_j-
 \zeta_i) e^{\frac{\im}{2}(\zeta_i^T\cdot J \cdot \zeta_j)} \geq 0
 \quad [\Rightarrow \quad \chi^*(\zeta,\eta) = \chi(-\zeta,-\eta)]
\end{equation}
for all $a_{i,j} \in \mathbb{R}$. This can be shown using the
following theorem.

\begin{theorem}
 \emph{(Quantum Bochner-Khinchin theorem)}
 \label{qbk}
 For $\chi(\eta)$ to be a characteristic function of a quantum
 state the following conditions are necessary and sufficient\\
 1.) $\chi(0) = 1$ and $\chi(\eta)$ is continuous at $\eta =
 0$,\\
 2.) $\chi(\eta)$ is $J-positive$ (symplectic-positive defined).
\end{theorem}

\subsection{Properties of the Wigner distribution}

{\em Properties}\\
{\em i) Quasidistribution:} It is real valued quasidistribution
because it admit negatives values (a Quantum Mechanics signature).\\
{\em ii) T-symmetry:} It has time symmetry
\begin{equation}
 t \rightarrow -t \Longleftrightarrow \mathcal{W}(q,p,t) \rightarrow
 \mathcal{W}(q,-p,t).
\end{equation}\\
{\em iii) X-symmetry:} It has space symmetry
\begin{equation}
 q \rightarrow -q \Longleftrightarrow
 \mathcal{W}(q,p,t) \rightarrow \mathcal{W}(-q,-p,t).
\end{equation}\\
{\em iv) Galilei invariant:} It is Galilei invariant
\begin{equation}
 q \rightarrow q-a \Longleftrightarrow
 \mathcal{W}(q,p,t) \rightarrow \mathcal{W}(q + a,p,t).
\end{equation}\\
{\em v) T-evolution:} The equation of motion for each point in the
phase space is classical in the absence of
forces~\footnote{Remember that when we are speaking about states
of light $m$ has to be interpret as permittivity of vacuum
$\epsilon_0$ while there is a minus sign difference with
Heisenberg's equation of motion.
\begin{equation}
 \frac{d\hat{A}}{dt} = \frac{1}{\im}\comm{A}{H}.
\end{equation}}
\begin{equation}
 \frac{d\hat{\rho}}{dt} = -\frac{1}{\im}\comm{\rho}{H}
 \Longleftrightarrow \frac{\partial \mathcal{W}(q,p,t)}{\partial
 t}
 = -\frac{p}{m} \frac{\partial \mathcal{W}(q,p,t)}{\partial q}.
\end{equation}\\
{\em vi) Bounded:} It is bounded~\footnote{Use Schwarz's inequality
$|\braket{\psi_1}{\psi_2}|^2 \leq \braket{\psi_1}{\psi_1}
\braket{\psi_2}{\psi_2}$ at the density operator level \em{i.e.}
$0 \leq \tr (\rho_1 \rho_2)^n \leq \tr (\rho_1)^n \tr (\rho_2)^n$. (no es ven be la dem)}
\begin{equation}\label{bounded}
 |\mathcal{W}(q,p)| \le \frac{1}{\pi}.
\end{equation}\\
{\em vii) Normalised:} It is well normalised
\begin{equation}
 \int dq \int dp \, \mathcal{W}(q,p) = 1.
\end{equation}\\
{\em viii) Quantum marginal distributions:} It possesses good
marginal distributions~\footnote{For pure state they correspond to
the square modulus of the wave function in position $|\psi(q)|^2$
and in momentum $|\tilde{\psi}(p)|^2$ representation.}
\begin{equation}
 \int dp \, \mathcal{W}(q,p) = \sandwich{q}{\rho}{q} \geq 0
\end{equation}
\begin{equation}
 \int dq \, \mathcal{W}(q,p) = \sandwich{p}{\rho}{p} \geq 0.
\end{equation}\\
{\em ix) Complete orthonormal set:} The set of functions
$\mathcal{W}_{nm}(q,p)$ form a complete orthonormal set (if
$\psi_{n}(q)$ are already a set)
\begin{equation}
 \int dq \int dp \, \mathcal{W}^*_{nm}(q,p)
 \mathcal{W}_{n'm'}(q,p) = \frac{1}{2 \pi} \, \delta_{nn'}
 \delta_{mm'}
\end{equation}
\begin{equation}
 \sum_{n,m} \mathcal{W}^*_{nm}(q,p) \mathcal{W}_{nm}(q',p') =
 \frac{1}{2 \pi} \delta(q-q') \delta(p-p')
\end{equation}
where
\begin{equation}
 \mathcal{W}_{nm}(q,p) = \frac{1}{\pi} \int dx \, e^{-2\im
 px} \psi^*_n(q-x) \psi_m(q+x).
\end{equation}

\subsection{The generating function of a Classical probability distribution}

Denoting by $y$ ($x$) a random variable which can be discrete $y
\in \{ y_i \}$ (or continuous $x \in [a,b]$) and its corresponding
(density) probability $p(y_i)$ ($p(x)$), we can establish the
normalisation constrain as

\begin{equation}
 \left. \begin{array}{lcr}
 \sum_{y_i}p(y_i) &=& 1\\
 \int_a^b p(x) dx &=& 1
 \end{array} \right\}.
\end{equation}

\noindent{Of relevant importance given a probability distribution
are the following quantities:}

\noindent{{\em i) Mean value of $u(x)$}: \quad $E[u(x)] = \int
u(x) p(x) dx$.\\}
{\em ii) Moment of order $m$ respect point $c$
of $x$}: \quad
$\alpha^m_c = E[(x-c)^m]$.\\
{\em iii) Mean value of $x$}: \quad $\mu = \alpha^1_0 = E[x]$.\\
{\em iv) Standard deviation of $x$:}~\footnote{Square root of the
variance.} \quad $\sigma = \sqrt{\alpha^2_\mu} =
\sqrt{E[(x-\mu)^2]}$.\\
{\em v) Covariance of $x_i$ and $x_j$:}~\footnote{Here subindex
$i,j$ labels all the possible variables of the distribution, when
they are equal, $C_{ii}$ corresponds to the variance of the
variable $x_i$.} \quad $C_{ij} = {\rm cov}(x_i,x_j) =
E[(x_i-\mu_i)(x_j-\mu_j)]$\\
where $i,j=1,2,...,2N$.

\begin{theorem}
 \emph{(Taylor's theorem)}
 \label{TT}
 Any well behaved distribution function can be reconstructed by its (in
 general) infinite moments.
\end{theorem}

This theorem, of considerable importance, tell us that any distribution
$p(x)$ can be retrieved only by its moments $\alpha^m_c$. We define
the vector $d$ and the matrix $C$ called mean vector and covariance
matrix by

\begin{equation}
 \left. \begin{array}{lcl}
 C &=& [[{\rm cov}(x_i,x_j)]]\\
 d &=& [[\mu_i]]
 \end{array} \right\}.
\end{equation}
What is more important is that $d$ and $C$ encode all the
information of $1^{\textrm{st}}$ and $2^{\textrm{nd}}$ moments.

If we define the generating function of the distribution function
by a Laplace transformation (provided it exists)
\begin{equation}
 M(\eta) = \mathcal {LT}\{ p(x) \} = E[e^{x \eta}]
\end{equation}
all moments can be obtained by subsequently differentiating the
generating function

\begin{equation}
 \alpha^m_0 = \frac{\partial^{(m)} M(\eta)}{\partial \eta^m}
 |_{\eta=0}.
\end{equation}

\subsection{The generating function of a quasi-probability distribution}

In the same way as in Classical Probability where all the moments
of a distribution characterise the distribution, the Wigner
quasidistribution function is fully characterised by its moments.

To adapt the classical formalism to the quantum Wigner
quasidistribution function we have to introduce the following
transcription $\eta \longrightarrow \im \eta, M \longrightarrow
\chi, {\mathcal LT} \longrightarrow {\mathcal FT}$.

We then define the generating function of the Wigner distribution
(characteristic function) by a Fourier transformation, which always
exists, because the Wigner distribution is an integrable function.
In general it is complex and reads

\begin{equation}
 \chi(\eta) = \mathcal {FT}\{ \mathcal{W}(x) \} = E[e^{\im x \eta}]
\end{equation}
then all moments can be obtained by subsequently differentiating
the generating function

\begin{equation}
 \beta^m_0 = \frac{1}{\im^m} \frac{\partial^{(m)} \chi(\eta)}
 {\partial \eta^m} |_{\eta=0}.
\end{equation}

Analogously, we define, given a quantum Wigner distribution
function the displacement vector (DV) $d$ (a $2N$ real vector) and
the covariance matrix (CM) $\gamma$ (a $2N \times 2N$ real
symmetric matrix). The DV contains the information of the first
moments and in general plays no role, by the space symmetry only
relative DVs are of physical meaning. The CM is much more richer,
it contains information (up to second moments) about the purity,
entanglement, ... The CM to describe a physical state must be
symplectic-positive defined

\begin{equation}
 \gamma + \im J \geq 0.
\end{equation}

\section{Gaussian states}

Among all the CV systems Gaussian states are of greatest
importance. The Gaussian distribution is simple, it is the limit of
many others and appears in a great variety of different conditions.
In order to give a motivation we state here one of the most important
theorems (together with the Law of Large Numbers) of the Theory
of the Classical Probability and Statistics, the Central Limit
Theorem.

\begin{theorem}
 \emph{(Central limit theorem)}
 \label{CLT}
 Suppose we have $n$ independent random variables $x_1, x_2,..., x_n$
 which are all distributed with a mean value $\mu$ and a standard
 deviation $\sigma$ (each of them can have different arbitrary
 distribution functions $p_i(x_i)$). In the limit $n \to \infty$ the
 arithmetic mean $\bar x = \frac{1}{n}\sum_{i=1}^n x_i$ is Gaussian
 (or normal) distributed with mean value $\mu$ and standard deviation
 $\frac{\sigma}{\sqrt{n}}$ {\em i.e.} $\bar p(\bar x) = \frac{1}{\sqrt{2
 \pi}\sigma}e^{-\frac{(\bar x - \mu)^2}{2 \sigma^2}}$.
\end{theorem}

Another way to see the importance of Gaussian probability
distributions is encoded in the following theorem

\begin{theorem}
 \emph{(Marcinkiewicz's theorem)}
 \label{MT}
 If we define the cumulant generating function as $K(\eta) =
 \ln{M(\eta)}$, then, either the cumulant is a polynomial of order 2
 or it has infinite terms.
\end{theorem}

\begin{lemma}
 \emph{(Gaussianity lemma)}
 \label{gauss}
 As a consequence then $p(x)$ is a Gaussian(non-Gaussian) distribution
 iff the cumulant is a polynomial of order 2($\infty$).
\end{lemma}

In general all moments are necessary but as long as we are concern
with only Gaussian distributions, $1^{\textrm{st}}$ and
$2^{\textrm{nd}}$ moments are sufficient, in fact all other higher
moments can be rewritten in terms of them. This is a consequence
of the theorem $\ref{MT}$.

\subsection{Displacement Vector (DV) and Covariance Matrix (CM)}

An important class of quantum states are Gaussian states. They can
be defined as all quantum states whose Wigner distribution
function is Gaussian. Thus the DV and the CM are enough to
describe them. Analogously to the classical case, it is
straightforward to obtain the moments of order $\beta^m_0$ of a
distribution by differentiating the generating function. Computing
$1^{\textrm{st}}$ and $2^{\textrm{nd}}$ moments, through
\eqref{characteristic} we get~\footnote{Notice: $(\frac{\partial}{\partial \eta_i} e^{\im \eta^T \cdot \hat{R}}) |_{\eta_i = 0} = \im \hat R_i$ and $(\frac{\partial^2}{\partial \eta_i \partial \eta_j} e^{\im \eta^T \cdot \hat{R}}) |_{\eta_{i,j} = 0} = \frac{1}{2} \comm{R_i}{R_j} - \hat R_i \hat R_j = -\frac{1}{2} \acomm{R_i}{R_j}$ where we have used Cambell-Hausdorff formula $e^{\hat A + \hat B} = e^{\hat A} e^{\hat B} e^{-\frac{1}{2} \comm{A}{B}}$ (when $\comm{A}{B} \propto \id$).}

\begin{equation}
 \beta^1_{0,i} = -\im \frac{\partial}{\partial \eta_i} \chi(\eta)
 |_{\eta = 0} = \tr (\hat{\rho} \hat{R'}_i)
\end{equation}

\begin{equation}
 \beta^2_{0,ij} = (-\im)^2
 \frac{\partial^2}{\partial \eta_i \partial \eta_j} \chi(\eta)
 |_{\eta = 0} = \frac{1}{2} \tr (\hat{\rho} \{
 \hat{R'}_i, \hat{R'}_j \}) = \tr (\hat{\rho} \hat{R'}_i
 \hat{R'}_j) - \frac{\im}{2} J_{ij}
\end{equation}
where $\hat{R'}_i = J \hat R_i$.

Finally we define the DV and the CM as~\footnote{For pure states
$d_i =  \exvalue{R_i}{\rho}$ and $\gamma_{ij} = \langle \{
\hat{R}_i-d_i \hat{\id}, \hat{R}_j-d_j \hat{\id} \} \rangle_{\rho}
= \langle \acomm{R_i}{R_j} \rangle_{\rho} - 2 \exvalue{R_i}{\rho}
\exvalue{R_j}{\rho}$, where, by the anticommutator definition, we
see a factor 2 of difference with the classical analog and so
$\gamma_{ij} \sim 2 C_{ij}$. The diagonal terms can be rewritten
in terms of the uncertainties as $\gamma_{ii} = 2 (\Delta
R_i)_\rho^2$ where as usual $(\Delta A)_\psi =
\sqrt{\exvalue{A^2}{\psi} - (\exvalue{A}{\psi})^2}$.}

\begin{equation}
 d_i = \tr (\hat{\rho} \hat{R}_i)
\end{equation}

\begin{equation}
\begin{split}
 \gamma_{ij} &= \tr (\hat{\rho} \{ \hat{R}_i-d_i \hat{\id},
 \hat{R}_j-d_j \hat{\id} \}) = 2 \tr [\hat{\rho}
 (\hat{R}_i-d_i \hat{\id}) (\hat{R}_j-d_j
 \hat{\id})] - \im J_{ij} =\\
 &= 2 {\rm Re} \{ \tr [\hat{\rho} (\hat{R}_i-d_i \hat{\id})
 (\hat{R}_j-d_j \hat{\id})] \}
\end{split}
\end{equation}

It is important to remark here that symplectic operations at the
level of the DV and CM act in such a way that any unitary
$\hat{U}_S$ maps to the following transformation $\gamma_S = S
\cdot \gamma \cdot S^T $ and $d_S = d + s$ where $S$ stands for an
element of the symplectic group, while $s$ stands for a phase space
translation.

With these definitions it can be shown that the Wigner distribution
of any Gaussian state can be written in terms of the DV and CM
through ~\footnote{We see here that from \eqref{bounded} ${\rm max}
\left[\mathcal{W}(\zeta)\right] = \mathcal{W}(d) = \frac{1}{\pi^N
\sqrt{\det \gamma}} \leq \frac{1}{\pi^N}$ where the equality holds
for pure states only.}

\begin{equation}\label{GWigner}
 \mathcal{W}(\zeta) = \frac{1}{\pi^N \sqrt{\det \gamma}}
 e^{-(\zeta-d)^T \cdot \frac{1}{\gamma} \cdot (\zeta -d)}
\end{equation}
while its symplectic-Fourier transform reads

\begin{equation}\label{GCharacteristic}
 \chi(\eta) = e^{\im \eta^T \cdot J \cdot d - \eta^T \cdot
 J^T\frac{\gamma}{4}J \cdot \eta} = e^{\im \eta^T
 \cdot d' - \eta^T \cdot \frac{\gamma'}{4} \cdot \eta}
\end{equation}
where $d'_i = J_{ij} d_j$ and $\gamma'_{ij} = J^T_{ik} \gamma_{kl}
J_{lj}$.

\begin{theorem}
 \emph{(Minimum uncertainty states theorem)}
 \label{mus}
 Equality in Heisenberg's uncertainty theorem is attained iff the
 state is a pure Gaussian state {\em i.e.} a rotated squeezed coherent
 state, $\ket{\psi} = \hat U_\theta \hat U_r \hat U_\alpha \ket{0}$.
\end{theorem}

All pure Gaussian states of one mode, characterised by its
$\gamma$ (and if necessary by $d$), can be obtained from the vacuum
state by an arbitrary displacement+squeezing+rotation in the phase
space. These states, by theorem $\ref{mus}$, are minimum
uncertainty states. Instead, mixed Gaussian states of
one mode can be all obtained from a thermal state by an arbitrary
displacement+squeezing+rotation.

As the cornerstone examples of Gaussian states, we have the
vacuum, coherent, squeezed and thermal states.

{\em *Vacuum}: $\ket{0}$

\begin{equation}
 \gamma_0 = \begin{pmatrix}
        1 & 0\cr
        0 & 1
       \end{pmatrix}, \quad d_0 = \begin{pmatrix}
        0\cr
        0
       \end{pmatrix}.
\end{equation}

{\em *(Pure) Coherent}:~\footnote{A Coherent state can
alternatively be defined as the eigenstate of the annihilation
operator, $\hat a \ket{\alpha} = \alpha \ket{\alpha}$. Coherent
states form an {\em overcomplete} non-orthogonal
($\braket{\alpha}{\alpha'} =
\exp{[-(|\alpha|^2+|\alpha'|^2)/2+\alpha^* \alpha']}$) set base
($\frac{1}{\pi}\int d^2\alpha \ketbra{\alpha}{\alpha}= \id$) of
vectors of the Hilbert space.} $\ket{\alpha} = \mathcal{\hat
D}(\alpha) \ket{0} = \hat U_\alpha \ket{0}$

\begin{equation}
 \gamma_\alpha = S_\alpha \gamma_0 S_\alpha^T = \begin{pmatrix}
        1 & 0\cr
        0 & 1
       \end{pmatrix}, \quad d_\alpha = d_0 + s_\alpha =
       \begin{pmatrix}
        q_0\cr
        p_0
       \end{pmatrix},
\end{equation}
where $\alpha = \alpha_R + \im \alpha_I = \frac{q_0 + \im p_0}
{\sqrt 2}$.

{\em *(Pure) Squeezed}: $\ket{r} = \mathcal{\hat S}(r) \ket{0} =
\hat U_r \ket{0}$

\begin{equation}
 \gamma_r = S_r \gamma_0 S_r^T = \begin{pmatrix}
        e^{-2r} & 0\cr
        0 & e^{2r}
       \end{pmatrix}, \quad d_r = d_0 + s_r = \begin{pmatrix}
        0\cr
        0
       \end{pmatrix}.
\end{equation}

{\em *(Mixed) Thermal}: $\hat \rho_\beta = \frac{1}{\pi M} \int
d^2\alpha \ketbra{\alpha}{\alpha} e^{-|\alpha|^2/M}$

\begin{equation}
 \gamma_\beta = \begin{pmatrix}
        2M+1 & 0\cr
        0 & 2M+1
       \end{pmatrix}, \quad d_\beta = \begin{pmatrix}
        0\cr
        0
       \end{pmatrix},
\end{equation}
where $M = \frac{1}{e^\beta - 1} \geq 0$ being $\beta$ the inverse
temperature.

\subsection{Hilbert space, phase space and DV$\&$CM connection}

We have already shown how to describe quantum states and
operations at the different levels {\em i.e.} Hilbert space, phase
space and DV$\&$CM. Two main connections are needed still to
perform calculations in the phase space: the ordering of the
operators and the metric between them.

The Weyl association rule tells us about the ordering operators.
Provided we are using with the Wigner distribution, which is
symmetrical ordered, when working with observables we have to take
into account that as we are in the phase space and we have avoided
its operator character we have to symmetrise them. The way we have
to symmetrise is

\begin{equation}
 e^{\im \zeta \hat{q} + \im \eta \hat{p}} \longrightarrow :e^{\im
 \zeta \hat{q} + \im \eta \hat{p}}: = e^{\im \zeta \hat{q} + \im
 \eta \hat{p}} \longleftrightarrow  e ^{\im \zeta q + \im \eta p},
\end{equation}
where $:\,\,\,:$ stands for the symmetrical order. In general for
a polynomial on $q$ and $p$

\begin{equation}
 \hat{q}^n\hat{p}^m \longrightarrow :\hat{q}^n\hat{p}^m: =
 \frac{1}{2^n}\sum_{r=0}^{n}\binom{n}{r}\hat{q}^r\hat{p}^m\hat{q}^{n-r} =
 \frac{1}{2^m}\sum_{r=0}^{m}\binom{m}{r}\hat{p}^r\hat{q}^n\hat{p}^{m-r}
 \longleftrightarrow q^np^m.
\end{equation}

As an example take the observable ${\mathcal QP}$, its quantum
associated operator is of course $\hat q \hat p$. We know that
$\hat q$ and $\hat p$ do not commute but in the phase space $qp$
and $pq$ are functionally treated in the same way. Imagine we need
to find its average value, we have then to remove the ambiguity by
totally symmetrising. The recipe is $\hat q \hat p \longrightarrow
:\hat{q}\hat{p}: = \frac{\hat{q}\hat{p}+\hat{p} \, \hat{q}}{2}
\longleftrightarrow qp$. And so the average to be performed is
\begin{equation}
 <{\mathcal QP}> = <\hat q \hat p >_\rho = <qp + \im/2>_{\mathcal W}
\end{equation}
because
\begin{equation}
 \hat{q} \hat{p} = \frac{\hat{q} \hat{p} +
 \hat{p} \hat{q}}{2} + \im/2.
\end{equation}

More important and relevant averages concern the moments which can be obtained {\em via} the Wigner distribution as

\begin{equation}
 d_i = \tr (\hat{\rho} \hat{R}_i)= \int d^{2N}\zeta \,\, \left[ \zeta_i \right] \mathcal{W}(\zeta)
\end{equation}

\begin{equation}
 \gamma_{ij} = \tr (\hat{\rho} \{ \hat{R}_i-d_i \hat{\id},
 \hat{R}_j-d_j \hat{\id} \}) = \int d^{2N}\zeta \,\, \left[ 2(\zeta_i - d_i)(\zeta_j - d_j) \right] \mathcal{W}(\zeta).
\end{equation}

\begin{theorem}
 \emph{(Quantum Parseval theorem)}
 \label{qpt}
 Let $\hat{W}_\zeta$ be a strongly continuous and irreducible Weyl
 system acting on the Hilbert space $\mathcal{H}_\Omega$ with phase
 space $\Omega$. Then $\hat A \mapsto \tr \{\hat A \hat{W}_\zeta \}$,
 with $\zeta \in\Omega$, is an isometric map from the Hilbert space
 $\mathcal{H}$ (Hilbert-Schmidt operators) onto the Hilbert space
 $\mathcal{L}^2(\Omega)$ (square-integrable measurable functions on
 $\Omega$) such that\\
\begin{equation}
 \tr (\hat A^\dag \hat B) = \frac{1}{(2\pi)^N} \int d^{2N}\zeta \,\,
 \tr \{\hat A \hat{W}_\zeta \}^*\tr \{\hat B  \hat{W}_\zeta \}.
\end{equation}
\end{theorem}

This theorem is of capital importance, it follows from it how to
compute the scalar product between operators

\begin{equation}
 \begin{split}
 \tr \{\hat{A}^{\dag} \hat{B}\} &= \frac{1}{(2\pi)^N} \int d^{2N}\eta \,
 \, {A^\chi}^*(\eta) B^\chi(\eta) =\\
 &= \frac{1}{(2\pi)^N} \int d^{2N}\zeta \, \,
 A^\mathcal{W}(\zeta) B^\mathcal{W}(\zeta),
 \end{split}
\end{equation}
the trace of an operator~\footnote{Use that $\id^\mathcal{W}=1$
and $\id^\chi=(2 \pi)^N \delta^{(2N)}(\eta)$ computed from eq.
\eqref{gwigner} and eq. \eqref{gcharacteristic}.}
\begin{equation}
 \tr \{\hat{A}\} = A^\chi(0,0) = \frac{1}{(2\pi)^N} \int d^{2N}\zeta \,
 \, A^\mathcal{W}(\zeta),
\end{equation}
and the expectation value of an observable
\begin{equation}
 \begin{split}
 \langle \hat{A} \rangle_{\rho} &= \tr
 \{\hat{\rho} \hat{A}\}
 = \frac{1}{(2 \pi)^N} \int d^{2N}\eta \, \, \chi^*(\eta) A^\chi(\eta)
 =\\
 &= \int d^{2N}\zeta \,\,\mathcal{W}(\zeta)A^\mathcal{W}(\zeta).
 \end{split}
\end{equation}
To justify the above expression we just need to define properly
the Fourier-Weyl transform as
\begin{equation}
 \begin{split}
 \hat{A} &= \mathcal{FWT} \{A^\chi(\eta)\} =\frac{1}{(2 \pi)^N} \int
 d^{2N}\eta \,\, A^\chi(\eta) \hat{W}_{-\eta} =\\
 &= \frac{1}{(2 \pi)^{2N}} \int d^{2N}\eta \int d^{2N}\zeta \,\,
 A^\mathcal{W}(\zeta) e^{\im \zeta^T \cdot J \cdot \eta} \hat{W}
 _{-\eta},
 \end{split}
\end{equation}
and its inverse

\begin{equation}\label{gcharacteristic}
 A^\chi(\eta) = \mathcal{FWT}^{-1} \{\hat{A}\} = \tr \{\hat{A}
 \hat{W}_{\eta} \},
\end{equation}

\begin{equation}\label{gwigner}
 A^\mathcal{W}(\bar\zeta,\bar\eta) = 2^N \int d^N\lambda \,
 \sandwich{\bar\zeta+\lambda}{A}{\bar\zeta-\lambda} e^{-2\im \bar\eta
 \lambda}
\end{equation}
where $\bar\zeta^T = (\zeta_1,\zeta_2,...,\zeta_N$) idem for
$\bar\eta$.

\begin{equation}
 A^\mathcal{W}(\zeta) = \mathcal {SFT} \{A^\mathcal{\chi}(\eta)
 \}~\footnote{Notice that $A^\mathcal{W} = (2\pi)^N\mathcal{W}$ if
 $\hat A = \hat\rho$ see eq. \eqref{wigner} (for normalisation
 convenience).} = \frac{1}{(2 \pi)^N} \int d^{2N}\eta \,\,
 A^\chi(\eta) e^{- \im \zeta^T \cdot J \cdot \eta},
\end{equation}

\begin{equation}
 A^\chi(\eta) = \mathcal {SFT}^{-1} \{A^\mathcal{W}(\zeta)\}
 ~\footnote{Notice that $A^\chi = \chi$ if $\hat A = \hat\rho$ see
 eq. \eqref{characteristic}.} = \frac{1}{(2 \pi)^N} \int d^{2N}\zeta
 \,\, A^\mathcal{W}(\zeta) e^{\im \zeta^T \cdot J \cdot \eta}.
\end{equation}

An important concept in Quantum Information is the fidelity
$\mathcal{F}$ between quantum states. The one we adopt here is
the so called Bures-Uhlmann fidelity and it is defined as follows

\begin{equation}
 \mathcal{F}(\hat{\rho}_1, \hat{\rho}_2) = \lbrack \tr
 \sqrt{\hat{\rho}_1^{1/2} \hat{\rho}_2 \hat{\rho}_1^{1/2}}
 \rbrack^2.
\end{equation}
It is symmetric and normalised between $1$ (equal states) and $0$
(orthogonal states). Its definition is simplified when one of
the two states is pure (say $\hat\rho_1$), in this case it
converges to the Hilbert-Schmidt fidelity

\begin{equation}
 \mathcal{F}(\hat{\rho}_1, \hat{\rho}_2) = \tr(\hat{\rho}_1
 \hat{\rho}_2) = \sandwich{\psi_1}{\rho_2}{\psi_1}.
\end{equation}

In case both states are pure, then, the fidelity becomes simply the
overlap between the two states

\begin{equation}
 \mathcal{F}(\hat{\rho}_1, \hat{\rho}_2) =
 |\braket{\psi_1}{\psi_2}|^2.
\end{equation}

It is useful here to use theorem $\ref{qpt}$ to evaluate the
Hilbert-Schmidt fidelity between two Gaussian state (at least when
one is pure)~\footnote{The second and third equality is true for
all CV states.}

\begin{equation}
 \begin{split}
 \mathcal{F} (\hat{\rho}_1, \hat{\rho}_2) &= \tr (\hat{\rho}_1
 \hat{\rho}_2) = \left(\frac{1}{2 \pi}\right)^N \int d^{2N}\eta \,
 \chi_1^*(\eta) \chi_2(\eta) = (2 \pi)^{N}
 \int d^{2N}\zeta \, \mathcal{W}_1(\zeta) \mathcal{W}_2(\zeta) =\\
 &= \frac{1}{\sqrt{\det
 (\frac{\gamma_1 + \gamma_2}{2})}} e^{-d^T (\frac{1}{\gamma_1 +
 \gamma_2}) d}
 \end{split}
\end{equation}
where $\gamma_{1(2)}$ and $d_{1(2)}$ belongs to
$\hat{\rho}_{1(2)}$, while $d=d_2-d_1$.\\

Another important concept in Quantum Information is the purity
$\mathcal{P}$ of a quantum state. The purity is defined as follows

\begin{equation}
 \mathcal{P}(\hat{\rho}) = \tr(\hat{\rho}^2)
\end{equation}

It is normalised between $1$ (pure states) and $0$ (maximally
mixed states). Also here using theorem $\ref{qpt}$ we can evaluate
the purity of a Gaussian state~\footnote{The first equality is
true for all CV states.}

\begin{equation}
 \mathcal{P}(\hat{\rho}) = (2 \pi)^{N} \int d^{2N}\zeta \, [\mathcal{W}
 (\zeta)]^2 = \frac{1}{\sqrt{\det \gamma}}
\end{equation}

\section{Multipartite states and entanglement}

If we want to treat the entanglement of a quantum state, first we
need to introduce multipartite states. In general we have to
extend the Hilbert space. At the level of density operators this
means that we have to ``tensor product`` $\otimes$, the Hilbert
space of each party {\em i.e.} $\mathcal{H} = \bigotimes_{k=1}^N
\mathcal{H}_k$. The covariance matrix structure for Gaussian
states turns to be simplified to a ''direct sum`` $\oplus$, of
each party's associated phase space {\em i.e.} $\Omega =
\bigoplus_{k=1}^N \Omega_k$. This is reminiscent of the Quantum
Parseval theorem, which transforms tensor product between density
matrices to products of Wigner functions (and Characteristic
functions) and at the same time direct sums of covariance
matrices.

Therefore, the advantage of using Gaussian states is that we fully
describe a state by a finite dimensional $2\times2$ matrix instead
of its infinite dimensional density matrix. Additionally,
dimensionality of the phase space increases slower, as dimensions
are added instead of multiplied.~\footnote{Remember that
$\dim(\hat\rho_1\otimes\hat\rho_2)=\dim(\hat\rho_1)\dim(\hat\rho_2)$
while
$\dim(\gamma_1\oplus\gamma_2)=\dim(\gamma_1)+\dim(\gamma_2)$.}

\subsection{Bipartite Gaussian states}

Any bipartite Gaussian state can be written in a block structure as
$\gamma = \begin{pmatrix}
    A & C\cr
    C^T & B
 \end{pmatrix}$, where $A=A^T$ and $B=B^T$.

\begin{lemma}
 \emph{(Normal form)}
 \label{nform}
 Every $1\times1$ mode Gaussian state can be transformed (by two
 local symplectic transformations) to\\
 \begin{equation}
 \gamma=\begin{pmatrix}
    \lambda_a & 0 & k_x & 0\cr
    0 & \lambda_a & 0 & k_p\cr
    k_x & 0 & \lambda_b & 0\cr
    0 & k_p & 0 & \lambda_b
 \end{pmatrix}.
 \end{equation}
\end{lemma}

If we define the four invariants of an arbitrary state $\mathcal A
= \det A$, $\mathcal B = \det B$, $\mathcal C = \det C$ and
$\Upsilon = \det \gamma$, then the following holds

\begin{equation}
 \begin{split}
 \lambda_a &= \sqrt \mathcal A\\
 \lambda_b &= \sqrt \mathcal B\\
 k_x &= \frac{1}{2}(\alpha-\sqrt{\alpha^2-4\mathcal C})\\
 k_p &= \frac{1}{2}(\alpha+\sqrt{\alpha^2-4\mathcal C})\\
 \alpha &= \sqrt{(\frac{(\sqrt{\mathcal A\mathcal B}+\mathcal C)^2
 -\Upsilon}{\sqrt{\mathcal A\mathcal B}})}
 \end{split}
\end{equation}

\begin{lemma}
 \emph{(Standard form)}
 \label{sform}
 Every $1\times1$ mode Gaussian state can be transformed (by local
 quasi-free symplectic transformations) to\\
 \begin{equation}\gamma=\begin{pmatrix}
    \lambda_a & 0 & k_x & 0\cr
    0 & \lambda_a & 0 & -k_p\cr
    k_x & 0 & \lambda_b & 0\cr
    0 & -k_p & 0 & \lambda_b
 \end{pmatrix},
\end{equation}
\end{lemma}
where $\lambda_a,\lambda_b \geq 1$ and $k_x \geq |k_p|$.

A Gaussian state in the Standard form is called symmetric if
$\lambda_a = \lambda_b$, and fully symmetric if it is symmetric and
in addition $k_x = k_p$.

\subsection{Entanglement of Gaussian states}\label{nega}

For discrete variable systems an important separability criteria
based on the partial transpose (time reversal) exists.

\begin{lemma}
 \emph{(NPPT Peres criteria)}
 \label{NPPTP}
 Given a bipartite state $\hat\rho$, if it has non-positive
 partial transpose ($\hat\rho^{T_A} \ngeq 0 \Rightarrow \hat\rho^{T_B}
 \ngeq 0$), then $\hat\rho$ is entangled.
\end{lemma}

\begin{lemma}
 \emph{(NPPT Horodecki criteria)}
 \label{NPPTH}
 In $\mathbb{C}^2\otimes\mathbb{C}^2$ and $\mathbb{C}^2\otimes
 \mathbb{C}^3$ given a bipartite state $\hat\rho$, it is entangled iff
 it has non-positive partial transpose ($\hat\rho^{T_A} \ngeq 0
 \Rightarrow \hat\rho^{T_B} \ngeq 0$).
\end{lemma}

For continuous variable states, Peres criteria also holds while
Horodecki criteria is true provided our state is composed of $1
\times N$ modes. In particular for Gaussian states, time reversal
is very easy to implement at the covariance matrix level. If $\hat
T$ is the reversal operator then $S_T =\theta= \begin{pmatrix}
    1 & 0\cr
    0 & -1
 \end{pmatrix}$ is the symplectic operations in phase space. So we
 can rewrite the lemma $\ref{NPPTH}$ for Gaussian states as

\begin{lemma}
 \emph{(NPPT Horodecki's criteria)}
 \label{GNPPTH}
 For $1 \times N$ modes given a bipartite Gaussian state $\gamma$,
 it is entangled iff it has non-positive partial transpose ($\theta_A
 \gamma \theta_A^T + \im J \ngeq 0 \Rightarrow \theta_B \gamma
 \theta_B^T + \im J \ngeq 0$).
\end{lemma}

Concerning entanglement measures we use as an entanglement measure
for pure state the Entropy of entanglement and for mixed ones the
Logarithmic negativity.\\

{\em *(Pure states)} Entropy of entanglement:
\begin{equation}
 E_S(\hat\rho) = S(\hat\rho_A) = -\tr(\hat\rho_A \log \hat\rho_A),
\end{equation}
where $S$ is the von Neumann Entropy~\footnote{The logarithm is in
base 2.} $S(\hat\rho) = -\tr(\hat\rho \log \hat\rho)$, and
$\hat\rho_A$ is the trace over $B$ defined as $\hat\rho_A = \tr_B
(\hat\rho)$. For any state it reduces (in terms of the Schmidt
coefficients) to
\begin{equation}
 E_S(\hat\rho) = -\sum_{i=0}^\infty C_i^2 \log C_i^2,
\end{equation}
while for Gaussian states,
\begin{equation}
 E_S(\gamma) = -\sum_{i=1}^{N_A}[(\frac{\mu_i+1}{2}) \log (\frac
 {\mu_i+1}{2}) - (\frac{\mu_i-1}{2}) \log (\frac{\mu_i-1}{2})],
\end{equation}
where $\{\mu_i\} = {\rm spec}(-\im J \gamma_A)$.\\

{\em *(Mixed states)} Logarithmic negativity (additive monotone):
\begin{equation}
 E_N(\hat\rho) = LN(\hat\rho)  = \log ||\hat\rho^{T_A}||_1,
\end{equation}
where $||\,\,||_1$ is the trace norm defined as $||\hat\rho||_1 =
\tr |\hat\rho| = \tr \sqrt{\hat\rho^T\hat\rho}=\sum {\rm singular
values} (\hat\rho)$. For Gaussian states,
\begin{equation}
 E_N(\gamma) = -\sum_{i=1}^N\log[\min(\tilde\mu_i,1)],
\end{equation}
where $\{\tilde \mu_i\} = {\rm spec}(-\im J \gamma^{T_A})$.\\

\section{Appendix of integrals}

Gaussian integrals:

\begin{equation}
 \int e^{-\zeta^T \cdot A \cdot \zeta} d^{2N}\zeta =
 \frac{\pi^N}{\sqrt{\det A}}
\end{equation}

\begin{equation}
 \int e^{-\zeta^T \cdot A \cdot \zeta - b^T \cdot \zeta}
 d^{2N}\zeta = \frac{\pi^N}{\sqrt{\det A}} e^{b^T \cdot
 \frac{1}{4A} \cdot b}
\end{equation}

\noindent{Non-Gaussian integrals:}

\begin{equation}
 \int \zeta_i^{m_i} \zeta_j^{m_j} \zeta_k^{m_k}{\cdot}{\cdot}{\cdot}
 e^{-\zeta^T \cdot A \cdot \zeta} d^{2N}\zeta = (-1)^{m_i + m_j + m_k
 + {\cdot}{\cdot}{\cdot}} \left. \frac{\pi^N}{\sqrt{\det A}} \frac
 {\partial^{m_i}}{\partial b_i^{m_i}} \frac{\partial^{m_j}}{\partial
 b_j^{m_j}} \frac{\partial^{m_k}}{\partial b_k^{m_k}} {\cdot}{\cdot}
 {\cdot} e^{b^T \cdot \frac{1}{4A} \cdot b} \right|_{b_i = 0, b_j = 0,
 b_k = 0, {\cdot}{\cdot}{\cdot}}
\end{equation}

\begin{equation}
 \begin{split}
 \int& \zeta_i^{m_i} \zeta_j^{m_j} \zeta_k^{m_k}{\cdot}{\cdot}{\cdot}
 e^{-\zeta^T \cdot A \cdot \zeta - c^T \cdot \zeta} d^{2N}\zeta =\\
 &= (-1)^{m_i + m_j + m_k + {\cdot}{\cdot}{\cdot}} \left. \frac{\pi^N}
 {\sqrt{\det A}} \frac{\partial^{m_i}}{\partial b_i^{m_i}} \frac{
 \partial^{m_j}}{\partial b_j^{m_j}} \frac{\partial^{m_k}}{\partial
 b_k^{m_k}} {\cdot}{\cdot}{\cdot} e^{(b + c)^T \cdot \frac{1}{4A}
 \cdot (b+ c)} \right|_{b_i = 0, b_j = 0, b_k = 0, {\cdot}{\cdot}
 {\cdot}}\\
 \end{split}
\end{equation}

\noindent{Useful integrals:}

\begin{equation}
 \int \mathcal{W}(\zeta) d^{2N} \zeta = 1
\end{equation}

\begin{equation}
 \int (\mathcal{W}(\zeta))^2 d^{2N} \zeta = \frac{1}{(2 \pi)^{N}
 \sqrt{\det \gamma}}
\end{equation}

\begin{equation}
 \int \chi(\eta) d^{2N} \eta = \frac{(4 \pi)^{N}}{\sqrt{\det
 \gamma}} e^{-d^T \frac{1}{\gamma} d}
\end{equation}

\begin{equation}
 \int (\chi(\eta))^2 d^{2N} \eta = \frac{(2 \pi)^{N}}{\sqrt{\det
 \gamma}} e^{-d^T \frac{4}{\gamma} d}
\end{equation}

\begin{equation}
 \int |\chi(\eta)|^2 d^{2N} \eta = \frac{(2 \pi)^{N}}{\sqrt{\det
 \gamma}}
\end{equation}

\chapter{Efficiency in QKD protocols with entangled Gaussian states}

Efficiency is a key issue in any real implementation of a
cryptographic protocol since the physical resources are not
unlimited. We will first show that Quantum Key Distribution is
possible with an ''Entanglement based`` scheme with NPPT symmetric
Gaussian states in spite of the fact that these systems cannot be
distilled with Gaussian operations (they are all bound entangled).
In this work we analyze the secrecy properties of Gaussian states
under Gaussian operations. Although such operations are useless
for quantum distillation, we prove that it is possible to distill
efficiently a secret key secure against finite coherent attacks
from sufficiently entangled Gaussian states with non-positive
partial transposition. Moreover, all such states allow for
efficient key distillation, when the eavesdropper is assumed to
perform individual attacks before the reconciliation process. In
section (\ref{QKD}) we present the academic protocol
\cite{Navascues05}, while in section (\ref{effQKD}) we present the
way to perform QKD with in the protocol in an efficient way.

\section{State of the problem: QKD with entangled Gaussian states}\label{QKD}

Before presenting the protocol it is important to notice that
Gaussian states always admit a purification. Thus, any mixed
Gaussian state of $N$ modes can be expressed as the reduction of a
pure Gaussian state of $2N$ modes of the form:
\begin{equation}\nonumber
    \gamma_{2N} =
    \begin{pmatrix}
    \gamma_{N} & C_N\\
    C_N^T & \theta_N \gamma_{N}\theta_N^T
    \end{pmatrix}, \quad \quad
    C_N = J_N \sqrt{-(J_N
    \gamma_N)^2-\id} \, \theta_N, \quad
    \quad \theta_N = \bigoplus_{i=1}^N
    \theta,
\end{equation}
such that the mixed state can be obtained after tracing out $N$
modes from $\gamma_{2N}$. Here $\theta
= \bigl(
\begin{smallmatrix}
    1 & 0\\
    0 & -1
\end{smallmatrix}
\bigr)$, which is the momentum reflection in phase-space, is the
associated symplectic operation.

For what follows it is also important the fact that any NPPT
Gaussian state can be mapped by Gaussian Local Operations and
Classical Communication (GLOCC) to an NPPT symmetric state of $1
\times 1$ modes {\em i.e.} preserving the amount of entanglement.

As the last remark, to deal with the content of the entanglement
in Gaussian states we are going to use the negativity. As it was
stated in section (\ref{nega}), in terms of CMs and for $1 \times
1$ and $1\times N$ modes of bipartite Gaussian states the PPT
criterion, which tells us that a state $\hat{\rho}$ is entangled
if and only if it has non positive partial transposition, reads
$\theta_A \gamma \theta_A^T + \im J < 0$.

With all the formalism at hand we now review the main steps of the
protocol used in \cite{Navascues05}. Without loosing generality,
and by virtue of the above properties of Gaussian states, one
should only consider the case in which Alice and Bob share many
copies of a quantum system of $1 \times 1$ symmetric NPPT Gaussian
state $\hat \rho_{AB}$. To extract a list of classically
correlated bits to establish a secret key, each party measures the
quadratures of her/his mode $\hat X_{A,B}$ and accepts only those
outputs $x_{A,B}$ for which both parties have a consistent result
$|x_A|=|x_B|=x_0$. With probability $p(i,j)$, each party
associates the classical bit $i=0(1)$ to her/his outcome
$+x_0(-x_0)$. The probability that their symbols do not coincide
is given by $\epsilon_{AB}=(\sum_{i\neq j}p(i,j)) /
(\sum_{i,j}p(i,j))$. Having fixed a string of $M$ classical
correlated values, they can apply Classical Advantage Distillation
\cite{Maurer93}. To this aim, Alice generates a random bit $b$ and
encodes her string of $M$ classical bits into a vector $\vec b$ of
length $M$ such that $b_{Ai}+b_i=b \hspace{-1.5mm} \mod (2)$. Bob checks that
for his symbols all results $b_{Bi}+b_i=b' \hspace{-1.5mm} \mod (2)$ are
consistent, and in this case accepts the bit $b$. The new error
probability is given by
\begin{equation}
    \epsilon_{AB,M} = \frac{(\epsilon_{AB})^M}{(1 - \epsilon_{AB}
    )^M+(\epsilon_{AB})^M} < \left( \frac{\epsilon_{AB}}{1 -
    \epsilon_{AB}} \right)^M,
\end{equation}
which tends to zero for sufficiently large $M$. The most general
scenario for eavesdropping is to assume that Eve has access to the
states before their distribution. Hence, the states that Alice and
Bob share correspond to the reduction of a pure 4-mode state. We
consider two types of attacks:
(i) individual (or incoherent) attack, where Eve performs
individual measurements, possibly non-Gaussian, over her set of
states and (ii) finite coherent (or collective) attack, where Eve
waits until the distribution has been performed, and, decides, which
collective measurement gives her more information on the final key.
Now, security with respect to individual attacks from the
eavesdropper Eve, can be established if
\begin{equation}
    \left(\frac{\epsilon_{AB}}{1 - \epsilon_{AB}}\right)^M <
    |\braket{e_{++}}{e_{--}}|^M,
\end{equation}
where $\ket{e_{\pm \pm}}$ denotes the state of Eve once Alice and
Bob have projected their states onto $\ket{\pm x_0}$. Notice that
Eve can gain information if the overlap between her states after
Alice and Bob have measured coincident results is sufficiently
small. The above inequalities come from the fact that in the case
of individual attacks the error on Eve's estimation of the final
bit $b$ is bound from below by a term proportional to
$|\braket{e_{++}} {e_{--}}|^M$ \cite{Navascues05}. Therefore,
Alice and Bob can establish a key if
\begin{equation}\label{Security}
    \frac{\epsilon_{AB}}{1 - \epsilon_{AB}} < |\braket{e_{++}}
    {e_{--}}|.
\end{equation}
In \cite{Navascues05} it was shown that any $1 \times 1$ NPPT
state fulfils the above inequality and thus any NPPT Gaussian
state can be used to establish a secure key in front of individual
eavesdropper attacks. If we assume that Eve performs more powerful
attacks, namely finite coherent attacks, then security is only
guaranteed if the much more restrictive condition
\begin{equation}\label{Security2}
    \frac{\epsilon_{AB}}{1 - \epsilon_{AB}} < |\braket{e_{++}}
    {e_{--}}|^2
\end{equation}
is fulfilled. This new inequality is violated by some NPPT states.
Notice that this implies that the analysed protocol is not
good for these states in this more general scenario. Nevertheless,
using the recent techniques of \cite{christandl04}, one can find
states for which the presented protocol allows to extract common
bits secure against this attack.

\section{Efficient QKD with entangled Gaussian states}\label{effQKD}

Let us now present our results. Notice that since security relies
on the fact that Alice and Bob have better correlations than the
information the eavesdropper can learn about their state, perfect
correlation is not a requirement to establish a secure key. We
denote Alice's outputs by $x_{0A}$  and we calculate which are the
outputs Bob can accept so that the correlation established between
Alice and Bob outputs can be used to extract a secret bit.

We use the standard form of a bipartite $1 \times 1$ mode Gaussian
state,
\begin{equation}
\gamma_{AB} = \begin{pmatrix}
    \lambda_A & 0 & c_x & 0\\
    0 & \lambda_A & 0 & -c_p\\
    c_x & 0 & \lambda_B & 0\\
    0 & -c_p & 0 & \lambda_B\\
    \end{pmatrix}
\end{equation}
with $\lambda_{A,B} \geq 1$, and $c_x \geq |c_p| \geq 0$ (we can
shift the displacement vector to 0). We shall deal with mixed
symmetric states and so $\lambda_A = \lambda_B = \lambda$. The
positivity condition reads $(\lambda - c_x)(\lambda + c_p) \geq
1$, while the entanglement NPPT condition is given by $(\lambda -
c_x)(\lambda - c_p) < 1$. As in \cite{Navascues05}, we impose that
the global state including Eve is pure (she has access to all
degrees of freedom outside Alice an Bob) while the mixed symmetric
state, shared by Alice and Bob is just its reduction,
\begin{equation}
    \gamma_{ABE} =
    \begin{pmatrix}
     \gamma_{AB} & C\\
     C^T & \theta \gamma_{AB} \theta^T
    \end{pmatrix},
\end{equation}
\begin{equation}
    C = J_{AB} \sqrt{-(J_{AB} \gamma_{AB})^2 - \id_2} \, \theta_{AB}
    = \begin{pmatrix}
     0 & -\textsc{X} & 0 & -\textsc{Y}\\
     -\textsc{X} & 0 & -\textsc{Y} & 0\\
     0 & -\textsc{Y} & 0 & -\textsc{X}\\
     -\textsc{Y} & 0 & -\textsc{X} & 0\\
    \end{pmatrix},
\end{equation}
\begin{equation}
    \theta_{AB} = \theta_A \oplus \theta_B, \quad \quad J_{AB} = J_A
    \oplus J_B,
\end{equation}
where
$$\textsc{X}=\frac{\sqrt{a+b} + \sqrt{a-b}}{2},$$
$$\textsc{Y}=\frac{\sqrt{a+b} - \sqrt{a-b}}{2},$$
and $a = \lambda^2 -c_x c_p - 1$, $b = \lambda(c_x - c_p)$.

Performing a measurement with uncertainty $\sigma$, the
probability that Alice finds $\pm |x_{0A}|$ while Bob finds $\pm
|x_{0B}|$, is given by the overlap between the state of Alice and
Bob, $\hat\rho_{AB}$, and a pure product state $\hat \rho_{A,i}
\otimes \hat \rho_{B,j}$ (with $i,j=0,1$) of Gaussians centred at
$\pm |x_{0A}|(\pm |x_{0B}|)$ respectively with $\sigma$ width
(notice $\hat\rho_{A,0} \equiv \ketbra{+|x_{0A}|}{+|x_{0A}|}$). We
use here the Hilbert-Schmidt fidelity for calculation, which
gives:
\begin{equation}\label{Prob00}
    \begin{split}
     p(0,0) &= p(1,1) = \tr[\hat \rho_{AB} (\hat \rho_{A,0}
     \otimes \hat \rho_{B,0})] =\\
     &= (2\pi)^4 \int d^4 \zeta_{AB} \, \mathcal{W}_{\rho_{AB}}
     (\zeta_{AB}) \mathcal{W}_{\rho_{A,0} \otimes
     \rho_{B,0}} (\zeta_{AB}) =\\
     &= K(\sigma) \exp \left( \frac{2|x_{0A}| |x_{0B}| c_x - (\lambda
     + \sigma^2)(x_{0A}^2 + x_{0B}^2)}{(\lambda + \sigma^2)^2 -
     c_x^2} \right),
    \end{split}
\end{equation}
for the probability that their symbols do coincide and,
\begin{equation}\label{Prob01}
    p(0,1) = p(1,0) = K(\sigma) \exp \left( \frac{-2|x_{0A}| |x_{0B}|
    c_x - (\lambda + \sigma^2)(x_{0A}^2 + x_{0B}^2)}{(\lambda +
    \sigma^2)^2 - c_x^2} \right),
\end{equation}
for the probability that they do not coincide, where
\begin{equation}
    K(\sigma) = \frac{4\sigma^2}{\sqrt{(\lambda + \sigma^2)^2 -
    c_x^2}\sqrt{(\lambda \sigma^2 + 1)^2 - c_p^2 \sigma^4}}.
\end{equation}
Their error probability for $\sigma \rightarrow 0$ reads
\begin{equation}\label{Errab}
    \epsilon_{AB} = \lim_{\sigma \to 0} \frac{\sum_{i \neq
    j}p\,(i,j)}{\sum_{i,j} p\,(i,j)} = \frac{1}{1 + \exp \left(
    \frac{4c_x |x_{0A}||x_{0B}|}{\lambda^2 - c_x^2} \right)}.
\end{equation}
Let us calculate the state of Eve $\ket{e_{\pm \pm}}$ after Alice
has projected onto $\ket{\pm |x_{0A}|}$ and Bob onto $\ket{\pm
|x_{0B}|}$:

\begin{equation}
    \gamma_{++} =  \gamma_{--} =
    \begin{pmatrix}
     \gamma_{x} & 0\\
     0 & \gamma_{x}^{-1}
    \end{pmatrix}, \quad \quad \gamma_{x} =
    \begin{pmatrix}
     \lambda & c_x\\
     c_x & \lambda
    \end{pmatrix},
\end{equation}

\begin{equation}
    d_{\pm\pm} = \mp
    \begin{pmatrix}
     0\\
     0\\
     A \delta x_0 - B \Delta x_0\\
     A \delta x_0 + B \Delta x_0
    \end{pmatrix},
\end{equation}
where $A = \frac{\sqrt{a+b}}{\lambda + c_x}$, $B
=\frac{\sqrt{a-b}}{\lambda - c_x}$, $\Delta x_0 = |x_{0B}| -
|x_{0A}|$ and $\delta x_0 = |x_{0B}| + |x_{0A}|$. The overlap
between the two states of Eve is given by:
\begin{multline}\label{Eveov}
    |\braket{e_{++}}{e_{--}}|^2 = \exp \Bigg(\frac{-4}{\lambda^2 -
    c_x^2} \Bigg[ \left( \frac{x_{0A}^2 + x_{0B}^2}{2} \right)
    (\lambda^2 - c_x^2-1)\lambda + \\ + |x_{0A}| |x_{0B}| \left(
    c_x - c_p(\lambda^2-c_x^2) \right) \Bigg] \Bigg).
\end{multline}
Substituting Eqs. \eqref{Errab} and \eqref{Eveov} into
\eqref{Security} one can check, after some algebra, that the last
inequality reduces to:
\begin{equation}\label{SecurityIneq}
    \left(\frac{x_{0A}^2 + x_{0B}^2}{2}\right)(\lambda^2 -
    c_x^2 - 1)\lambda + |x_{0A}| |x_{0B}| \left( -c_x - c_p
    (\lambda^2 - c_x^2) \right) < 0.
\end{equation}
Notice that condition \eqref{SecurityIneq} imposes both,
restrictions on the parameters defining the state ($\lambda, c_x,
c_p$), and on the outcomes  of the  measurements ($x_{0A},
x_{0B}$).  The constraints on the state parameters are equivalent
to demand that the state is NPPT and satisfies
\begin{equation}\label{Constrain}
    (\lambda - c_x)(\lambda + c_x) \geq 1.
\end{equation}
Nevertheless, as $c_x \ge c_p$, any positive state fulfils this
condition. Hence for any NPPT symmetric state, there exists, for a
given $x_{0A}$, a range of values of $x_{0B}$ such that secret
bits can be extracted (Eq. \eqref{Security} is fulfilled). This
range is given by
\begin{equation}
    \Delta x_0 = |x_{0B}| - |x_{0A}| \in {\mathfrak D}_\alpha =
    \left[ \frac{2}{-\sqrt \alpha-1} , \frac{2}{\sqrt \alpha-1}
    \right] |x_{0A}|,
\end{equation}
where
\begin{equation}
    \alpha = \left( \frac{c_x -
    \lambda}{c_x + \lambda} \right) \left[ \frac{1 - (\lambda +
    c_x)(\lambda + c_p)}{1- (\lambda - c_x)(\lambda - c_p)}
    \right].
\end{equation}
After Alice communicates $|x_{0A}|$ to Bob, he will accept only
measurement outputs within the above interval. The interval
$\Delta x_0$ is well defined if $\alpha \geq 1$, which equals to
fulfil Eq. \eqref{Constrain}. Notice also that the interval is not
symmetric around $|x_{0A}|$ because the probabilities calculated
in Eqs. \eqref{Prob00} and \eqref{Prob01} do depend on this value
in a non-symmetric way. The length $D_\alpha$ of the interval of
valid measurements outputs for Bob is given by
\begin{equation}
    D_\alpha = \frac{4 \sqrt \alpha}{\alpha-1} |x_{0A}|.
\end{equation}
It can be observed that maximal $D_\alpha\rightarrow\infty$
($\alpha=1$) corresponds to the case when Alice and Bob share a
pure state (Eve is disentangled from the system) and thus
condition \eqref{Security} is always fulfilled.  On the other
hand, any mixed NPPT symmetric state ($\alpha > 1$) admits a
finite $D_\alpha$. This ensures a {\em finite} efficiency on
establishing a secure secret key in front of individual attacks.

If we assume that Eve performs more powerful attacks, namely
finite coherent attacks, then security is only guaranteed if
\cite{Navascues05}:
\begin{equation}
    \frac{\epsilon_{AB}}{1 - \epsilon_{AB}} < |\braket{e_{++}}
    {e_{--}}|^2.
\end{equation}
This condition is more restrictive than \eqref{Security}. With a
similar calculation as before we obtain that now security is not
guaranteed for all mixed entangled symmetric NPPT states, but only
for those that also satisfy:
\begin{equation}\label{Constrain2}
    \lambda - (\lambda+c_x)(\lambda-c_x)(\lambda-c_p) > 0.
\end{equation}
For such states, and given a measurement result $x_{0A}$ of Alice,
Bob will only accept outputs within the range:
\begin{equation}
    \Delta x_0 = |x_{0B}| - |x_{0A}| \in {\mathfrak D}_\beta =
    \left[ \frac{2}{-\sqrt \beta-1} , \frac{2}{\sqrt \beta-1}
    \right] |x_{0A}|,
\end{equation}
where
\begin{equation}
    \beta = \frac{2\lambda(\lambda^2-c_x^2-1)}{\lambda-
    (\lambda+c_x)(\lambda-c_x)(\lambda-c_p)} \geq 1.
\end{equation}
As before, $\beta \geq 1$ is fulfilled by conditions
\eqref{Constrain} and \eqref{Constrain2}.

Let us now focus on the efficiency issue. We define the efficiency
$E(\gamma_{AB})$ of the protocol for a given state $\gamma_{AB}$,
as the average probability of obtaining a classically correlated
bit. Explicitly,
\begin{equation}\label{protocolefficiency}
    E(\gamma_{AB}) = \int_{\Delta x_0  \in {\mathfrak D}} dx_{0A}
    dx_{0B}(1-\epsilon_{AB}) \tr (\hat\rho_{AB} \ketbra{x_{0A},
    x_{0B}}{x_{0A},x_{0B}}).
\end{equation}
The marginal distribution in phase-space is easily computed by
integrating the corresponding Wigner function in momentum space
\cite{Lee95}:
\begin{equation}
    \begin{split}
     \tr (\hat\rho_{AB} \ketbra{x_{0A},x_{0B}}{x_{0A},x_{0B}})
     &= \int \int dp_A dp_B \mathcal{W}_{\rho_{AB}} ({\zeta}_{AB})
     =\\ &= \frac{\exp \left( \frac{2 c_x x_{0A} x_{0B}
     - \lambda (x_{0A}^2+x_{0B}^2)}{\lambda^2-c_x^2} \right) }
     {\pi \sqrt{\lambda^2-c_x^2}},
    \end{split}
\end{equation}
but the final expression of Eq. \eqref{protocolefficiency} has to be
calculated numerically. Note that if Alice and Bob share as a
resource $M$ identical states (NPPT state for individual attacks,
and NPPT fulfilling condition \eqref{Constrain2} for finite coherent
attacks), the number of classically correlated bits that can be
extracted from them is $\sim~M \times E(\gamma_{AB})$. The
efficiency Eq. \eqref{protocolefficiency} increases with increasing
$D$ and decreasing $\epsilon_{AB}$. In particular, for the
protocol given in \cite{Navascues05}, $D=0$, and therefore
$E(\gamma_{AB})=0$ for any state.

We investigate now the dependence of $E(\gamma_{AB})$ with the
entanglement of the NPPT mixed symmetric state used for the
protocol as well as with the purity of the state. As a measure of
the entanglement between Alice and Bob we compute the logarithmic
negativity
\begin{equation}
    {\rm LN} (\gamma_{AB}) = \log_2 \left( \frac{1}
    {\sqrt{(\lambda-c_x)(\lambda-c_p)}} \right) > 0.
\end{equation}
\begin{figure}
    \centering
    \includegraphics[width=10cm]{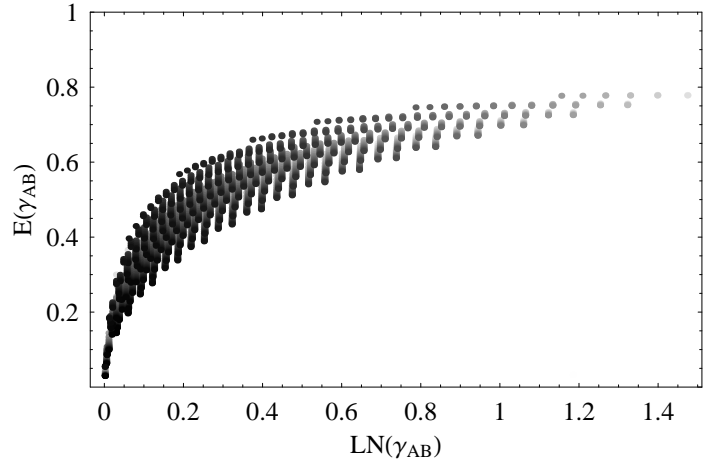}
    \caption{Protocol efficiency (quantified by $E(\gamma_{AB})$)
    versus the entanglement measured by logarithmic negativity
    ${\rm LN}(\gamma_{AB}$). The shading from black to white
    corresponds to purity from zero to one.}\label{fig1}
\end{figure}
In Fig. \ref{fig1}, we display the efficiency of the protocol
(assuming individual attacks) versus entanglement shared between
Alice and Bob for different states $\gamma_{AB}$. There is not a
one-to-one correspondence between $E(\gamma_{AB})$ and
entanglement, since states with the same entanglement can have
different purity, which can lead to different efficiency. This is
so because there are two favourable scenarios to fulfil Eq.
\eqref{Security}. The first one is to demand large correlations so
that the relative error $\epsilon_{AB}$ of Alice and Bob is small.
The second scenario happens when Alice and Bob share a state with
high purity, {\em  i.e.}, Eve is very disentangled. In this case,
independently of the error $\epsilon_{AB}$, Eq. \eqref{Security}
can be fulfilled more easily.

Despite the fact that efficiency generally increases with
increasing entanglement, this enhancement, as depicted in the
figure, is a complex function of the parameters involved.
Nevertheless, one can see that there exist an entanglement
threshold (around ${\rm LN}(\gamma_{AB}) \simeq 0.2$) below which
the protocol efficiency diminishes drastically no matter how mixed
are the states shared between Alice and Bob.

It is also illustrative to examine the dependence of $\alpha$
(which determines the interval length $D_\alpha$) on the
entanglement of the states shared by Alice and Bob.
\begin{figure}
    \centering
    \includegraphics[width=10cm]{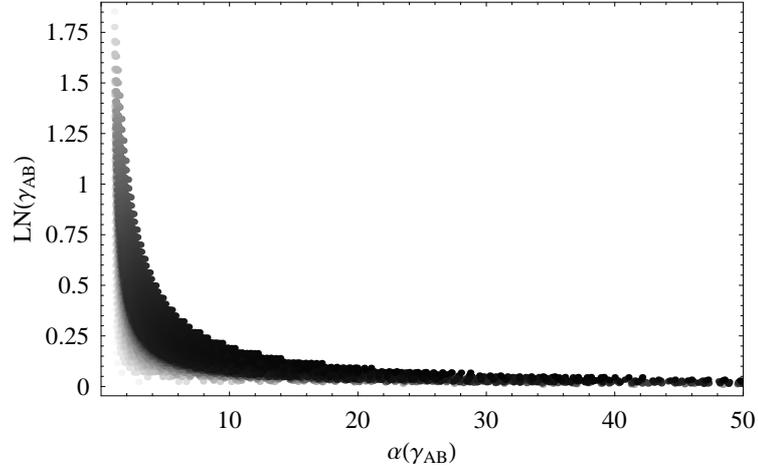}
    \caption{Entanglement of the states shared between Alice and Bob
    measured in terms of the logarithmic negativity ${\rm
    LN}(\gamma_{AB})$ as a function of the parameter $\alpha
    (\gamma_{AB})$ under individual attacks. The shading from black
    to white corresponds to purity from zero to one.}\label{fig2}
\end{figure}
In Fig. \ref{fig2} we plot the logarithmic negativity of a given
state versus the parameter $\alpha$. States with the same
entanglement but different purity are associated to quite
different values of $\alpha$. Nevertheless states with high
entanglement permit a large interval length (small $\alpha$) and,
thus, high efficiency.

In both, Fig. \ref{fig1} and Fig. \ref{fig2}, we have observed
that states with different entanglement give the same efficiency.
However it is important to pointed out that to extract the key's
bits, Classical Advantage Distillation \cite{Maurer93} stills
needs to be performed. The efficiency of Maurer's protocol,
strongly increases with decreasing $\epsilon_{AB}$, and,
therefore, the states with higher entanglement will provide a
higher key rate.

\chapter{Summary and Conclusions}

Efficiency is a key issue for any experimental implementation of
Quantum Cryptography since available resources are not unlimited.
Here, we have shown that the sharing of entangled Gaussian
variables and the use of only Gaussian operations permits
efficient Quantum Key Distribution against individual and finite
coherent attacks.

We have used the fact that all mixed NPPT symmetric states can be
used to extract secret bits under individual attacks whereas under
finite coherent attacks an additional condition has to be fulfilled.
We have introduced a figure of merit (the efficiency $E$) to
quantify the number of classical correlated bits that can be used to
distill a key from a sample of $M$ entangled states. We have observed that this
quantity grows with the entanglement shared between Alice and Bob.
This relation it is not one-to-one due to the fact that states
with less entanglement but with more purity (eavesdropper more
disentangled) can be equally efficient. Nevertheless we have point
out that, these states would be inefficient, when performing the
Classical Advantage Distillation of the key.

Finally, we would like to remark that our study is not restricted
to Quantum Key Distribution protocols, but can be extended to any
other protocol that uses as a resource entangled continuous
variables to establish a set of classically correlated bits
between distant parties, see {\em e.g.} \cite{Rodo08Bis}.

In \cite{Rodo08Bis} an efficient solution of the Byzantine
Agreement problem (detectable broadcast) in the continuous
variable scenario with multipartite entangled Gaussian states and
Gaussian operations (homodyne detection) is presented. In a
cryptographic context, detectable broadcast refers to distributed
protocols in which some of the participants might have malicious
intentions and could try to sabotage the distributed protocol
inducing the honest parties to take contradictory actions between
them. Entanglement is used in the protocol to distribute classical
private random variables with a specific correlation between the
players, in such a way that any malicious manipulation of the data
can be detected by all honest parties allowing them to abort the
protocol. We discuss realistic implementations of the protocol,
which consider the possibility of having inefficient homodyne
detectors, not perfectly correlated outcomes, and noise in the
preparation of the resource states. The proposed protocol is
proven to be robust and efficiently applicable under such
non-ideal conditions.

Following \cite{Wolf06}, it is known that, in spite of their
exceptional role within the space of all continuous variables states,
in fact, Gaussian states are not the best candidates as resources to
perform Quantum tasks. In general, any continuous, strongly,
super-additive functional acting on any given covariance matrix is
minimised by Gaussian states. This is the case for all entanglement
measures fulfiling the above conditions {\em e.g.} the distillable
entanglement or the entanglement of formation. In this sense Gaussian
states are extremal. With this idea on mind, naturally one could try
to perform QKD with non-Gaussian states. Following the presented
protocol here one should expect an enhancement on the efficiency on
the key distribution with non-Gaussian states. Gaussian states possess
an easy mathematical description at the covariance matrix level while
for non-Gaussian states this description is not complete.
Nevertheless, the Wigner distribution function formalism presented
here, allows to perform the needed calculations to study QKD with
non-Gaussian states, in a very similar way.

Also, for non-Gaussian states there are no computable entanglement
measures while these are well established in the case of Gaussian
states. Thus, one might think that a way to quantify entanglement in
non-Gaussian states can be accomplished by relating the efficiency
of distilling correlated bits with the entanglement of non-Gaussian
states \cite{Rodo08}. From the experimental point of view, there already exist several
groups which actually have succeeded in producing non-Gaussian states
like photon-substracted, states that up to now lack of a complete well
caracterisation.

\bibliography{Tesina}

\addcontentsline{toc}{chapter}{Bibliography}
\bibliographystyle{unsrt}

\end{document}